\documentstyle[aaspp4,epsf]{article}

\begin{document}

\title{Galaxy Clustering and Large--Scale Structure \\ from $z = 0.2$ to
$z = 0.5$ in Two Norris Redshift Surveys}

\author{Todd A. Small\altaffilmark{1}}
\affil{Institute of Astronomy, University of Cambridge,
Madingley Road, Cambridge CB3 0HA, UK \\
Electronic mail: tas@astro.caltech.edu}

\author{Chung-Pei Ma}
\affil{Dept.~of Physics and Astronomy,
University of Pennsylvania, Philadelphia, PA 19104 \\
Electronic mail: cpma@dept.physics.upenn.edu}

\author{Wallace L.W. Sargent}
\affil{Palomar Observatory, California Institute of
Technology, Pasadena, CA  91125 \\
Electronic mail: wws@astro.caltech.edu}

\author{and Donald Hamilton}
\affil{Institute of Astronomy and Astrophysics, University of
Munich, Scheinerstrasse 1, D-81679, Munich, Germany \\
Electronic mail: ham@usm.uni-muenchen.de}

\altaffiltext{1}{present address:  Space Astrophysics, 405-47,
California Institute of Technology, Pasadena, CA  91125}

\begin{abstract}
We present a study of the nature and evolution of large--scale
structure based on two independent redshift surveys of faint field
galaxies conducted with the 176--fiber Norris Spectrograph on the
Palomar 200--inch telescope.  The two surveys sparsely cover $\sim 20$
sq.\ degrees and contain 835 $r \le 21$ mag galaxies with redshifts
$0.2 < z < 0.5$.  Both surveys have a median redshift of $z_{\rm med}
\approx 0.30$.  In order to obtain a rough estimate of the cosmic
variance, we analyze the two surveys independently.

We have measured the two--point spatial correlation function and the
pairwise velocity dispersion for galaxies with $0.2 < z < 0.5$.  We
measure the comoving correlation length to be $3.70 \pm 0.13 h^{-1}$
Mpc ($q_0 = 0.5$) at $z_{\rm med} = 0.30$ with a power--law slope
$\gamma = 1.77 \pm 0.05$.  Dividing the sample into low ($0.2 < z <
0.3$) and high ($0.3 < z < 0.5$) redshift intervals, we find no
evidence for a change in the comoving correlation length over the
redshift range $0.2 < z < 0.5$.  Similar to the well--established
results in the local universe, we find that intrinsically bright
galaxies are more strongly clustered than intrinsically faint galaxies
and that galaxies with little ongoing star formation, as judged from
the rest--frame equivalent width of the [\ion{O}{2}]$\lambda$3727, are
more strongly clustered than galaxies with significant ongoing star
formation.  The rest--frame pairwise velocity dispersion of the sample
is $326^{+67}_{-52}$ km s$^{-1}$, $\sim 25$\% lower than typical
values measured locally.  Our sample is still too small to obtain
useful constraints on mean flows.

The appearance of the galaxy distribution, particularly in the more
densely sampled Abell 104 field, is quite striking.  The pattern of
sheets and voids which has been observed locally continues at least to
$z \lesssim 0.5$.  A friends--of--friends analysis of the galaxy
distribution supports the visual impression that $\gtrsim90$\% of all
galaxies at $z \lesssim 0.5$ are part of larger structures with
overdensities of $\gtrsim 5$, although these numbers are
sensitive to the precise parameters chosen for the
friends--of--friends algorithm.

\end{abstract}

\section{Introduction}

The clustering of galaxies is customarily characterized by a hierarchy
of n--point correlation functions (Peebles 1980), although, in
practice, only the two--point correlation function, $\xi(r)$, can be
measured accurately with current redshift surveys.  For local,
optically--selected samples, $\xi(r)$ is well--described by a power
law, $\xi(r) = (r/r_0)^{-\gamma}$, with correlation length $r_0
\approx 5 h^{-1}$ Mpc ($h$ is the present value of the Hubble constant
measured in units of 100 km s$^{-1}$ Mpc$^{-1}$) and slope $\gamma
\approx 1.8$ for $r \lesssim 20 h^{-1}$ Mpc (Loveday et al.\ 1995;
Marzke et al.\ 1995; Guzzo et al.\ 1997; Jing, Mo, \& B\"{o}rner 1998).  At
higher redshifts, the strength of the clustering can be inferred from
the two--point angular correlation function or measured directly from
a redshift survey.  Angular correlation function studies are essential
for understanding the clustering of faint objects beyond the
spectroscopic reach of current facilities and for reducing cosmic
variance by covering large areas (Brainerd \& Smail 1998, Postman et
al.\ 1998).  However, such studies must rely on models of the redshift
distribution of faint galaxies to deduce the three--dimensional
correlation length.

With the recent completion of large redshift surveys of galaxies with
redshifts up to $z \sim 1$ (e.g., Lilly et al.\ 1995a [CFRS]; Ellis et
al.\ 1996 [Autofib]; Cowie et al. 1996 [Hawaii]; Yee, Ellingson, \&
Carlberg 1996 [CNOCI]; Small, Sargent, \& Hamilton 1997a [Norris];
Carlberg et al.\ 1999 [CNOCII]), it is now possible to measure the
evolution of clustering directly.  Having been principally designed to
study the redshift--evolution of the galaxy population (to which they
have made landmark contributions), the CFRS, Autofib, and Hawaii
surveys are not particularly well--suited to studying large--scale
structure, mainly because they do not extend over large contiguous
areas of the sky.  Using the Autofib survey, Cole et al.\ (1994)
measured $\xi(r)$ for $z \lesssim 0.3$ and found no evidence for
evolution in the comoving correlation length.  In contrast, Le~F\`evre
et al.\ (1996) observed rapid decline of the correlation length with
redshift for galaxies observed in the Canada--France Redshift Survey.
Shepherd et al.\ (1997) used data from the CNOCI cluster survey to
measure a correlation length at $z \sim 1/3$ consistent with the
evolution inferred by Le~F\`evre et al.\ (1996).  The correlation
length of galaxies from the Hawaii K--band--selected sample exhibits
a similar decline over a large redshift range from $z \sim 0.3$ to $z
\sim 1.4$ (Carlberg et al.\ 1997).

The differing results for the redshift--evolution of galaxy clustering
reached so far emphasize the need for larger surveys to map
large--scale structure and to limit the impact of field--to--field
variance.  Indeed, Postman et al.\ (1998) from their I--band imaging
survey of 16 sq.\ deg.\ infer a correlation length at $z = 0.5$ twice
as large as that found, for example, in the CFRS.  Our two Norris
surveys of the fields of the Corona Borealis supercluster and the
Abell 104 galaxy cluster, while not as deep as the CFRS, Autofib,
Hawaii, and CNOC surveys, cover (albeit sparsely) fields of 16
sq. degrees and 4 sq. degrees, respectively, and include a total
comoving volume between $0.2 < z < 0.5$ of $2 \times 10^6 h^{-3}$
Mpc$^3$ ($q_0 = 0.5$).  Both Norris surveys, with the foreground
supercluster and cluster regions removed, have median redshifts of
0.3, and each contains several hundred $r \le 21$ mag galaxies with
redshifts in the interval $0.2 < z < 0.5$.  We expect any overall
density fluctuation,
\begin{equation}
{\delta\bar n \over \bar n} \sim {J_3 \overwithdelims() V}^{1/2}
\end{equation}
(Davis \& Huchra 1982), where $J_3$ is the second moment of the
two--point spatial correlation function (Peebles 1980) and is
approximately equal to $10,000 h^{-1}$ Mpc$^3$ (Tucker et al.\ 1997),
in our survey volume $V$ to be $\lesssim 10$\%.  This is comparable to
the statistical error in our estimates of the correlation strength at
$1 h^{-1}$ Mpc.  The only survey with comparable power for exploring
large--scale structure at intermediate--redshifts is the
recently--completed CNOCII survey (Carlberg et al.\ 1999).  The CNOCII
survey contains five times as many galaxies as our two surveys
combined but covers only 10\% of the area (in four separate survey
zones).  The preliminary results of the CNOCII survey presented by
Carlberg et al. (1999) are in general agreement with and complementary
to the main conclusions presented in this paper.

Observations of galaxy clustering such as those described here cannot,
however, be interpreted in a straightforward fashion because one may
not be observing the same types of galaxies at high redshift as at low
redshift.  In local samples, early-type galaxies are clustered more
strongly than late-type galaxies (Loveday et al.\ 1995, Guzzo et al.\
1997).  The importance of accounting for the change in the observed
galaxy population is illustrated by the strong clustering exhibited by
the Lyman--break galaxies at $z \sim 3$ (Steidel et al.\ 1997,
Giavalisco et al.\ 1998).  This strong clustering, with a correlation
length at least twice as large as that observed at $z \sim 1$ by
Le~F\`evre et al.\ (1996), can be naturally explained if the
Lyman--break galaxies are highly biased with respect to the mass
distribution (Bagla 1998, Steidel et al.\ 1998).  The likely variation
of bias with redshift and among different galaxy populations at the
same redshift will complicate efforts to test the gravitational
instability hypothesis and to determine the mass density of the
universe, but it also implies that studies of the growth of clustering can
help us to understand bias and galaxy formation.

The observed two--point correlation function, which is expected to be
isotropic in real space when averaged over a sufficiently large
volume, is distorted in redshift space by line--of--sight peculiar
velocities.  (For a comprehensive review, see Hamilton 1998.)  On
small scales, the velocity dispersion of bound clusters of galaxies
suppresses the apparent correlation function, whereas coherent motions
of galaxies towards overdense regions and away from underdense regions
enhance the correlation function on large scales.  An analysis of the
distortions allows one in principle to measure the moments of the
distribution of pairwise velocity distribution.  As is well known,
both the first and second moments can be used to estimate the mean
density of the universe, $\Omega_0$ (modulo the bias parameter).
However, we will not attempt to do so here since survey volumes
substantially larger than ours are required to obtain interesting
results (Fisher et al.\ 1994).

Our data are, nevertheless, well--suited to a measurement of the
pairwise velocity dispersion of galaxies, $\sigma_{12}$.  The pairwise
velocity dispersion is a measure of the kinetic energy in the galaxy
distribution.  Since $\sigma_{12}$ is a pair--weighted statistic and
thus very sensitive to the number and treatment of rich clusters in
the survey volume, it is difficult to compare values of $\sigma_{12}$
measured in different redshift surveys and with $\sigma_{12}$
determined in large $N$--body simulations in a consistent fashion.
The first measurement of the pairwise velocity dispersion,
$\sigma_{12} = 340 \pm 40$ km s$^{-1}$, was obtained by Davis \&
Peebles (1983) using data from the CfA1 redshift survey.  Subsequent
measurements, including a reanalysis of the CfA1 data by Somerville,
Davis, \& Primack (1997), have demonstrated as much as a factor of 2
variation in $\sigma_{12}$ depending on the type and environment of
galaxies analyzed and on the treatment of rich clusters (e.g., Mo,
Jing, \& B\"{o}rner 1993; Zurek et al.\ 1994, Marzke et al.\ 1995;
Guzzo et al.\ 1997; Jing, Mo, \& B\"{o}rner 1998).  However, redshift
surveys are becoming large enough and analysis techniques
sophisticated enough to begin to measure reliable values of
$\sigma_{12}$.  The large volume of our combined surveys, $\sim 2
\times 10^6 h^{-3}$ Mpc$^3$ (comoving, $q_0 = 0.5$), will enable us to
make a measurement of $\sigma_{12}$ at intermediate redshift for which
the error due to cosmic variance will be $\lesssim 20$\% (Marzke et
al.\ 1995), similar in size, in fact, to our statistical error.

We describe our data in the following section, \S 2.  The techniques
we use to compute the correlation function are outlined in \S 3, and
we present the results of this analysis in \S 4.  We discuss the
pairwise velocity dispersion $\sigma_{12}$ at $z_{\rm med} = 0.30$ in
\S 5.  In \S 6, we consider the large--scale structure of the galaxy
distribution in our sample and quantify the clustering on 10 to 100
$h^{-1}$ Mpc scales with a friends--of--friends analysis.  Finally, we
summarize our results and discuss the redshift evolution of $\xi$ and
$\sigma_{12}$ in \S 7.  We use $q_0 = 0.5$ in the main discussion and
for the figures except where explicitly noted.

\section{Data}

Our analysis is based on data we have obtained with the 176--fiber
Norris Spectrograph (Hamilton et al.\ 1993) on the Palomar 5 m
telescope.  We have surveyed two fields, one centered on the $z
\approx 0.07$ Corona Borealis supercluster (R.A. = $15^h30^m$, Dec. =
$+30^\circ$) and the other on the $z \approx 0.08$ Abell 104 galaxy
cluster (R.A. = $0^h45^m$, Dec. = $+24^\circ$).  The Norris
Spectrograph is designed for redshift surveys of faint galaxies.  Its
fibers are only 2 m long in order to minimize light losses due to
absorption in the fibers, and the fiber entrance aperture is only 1.6
arcsec (FWHM), which maximizes the contrast of a $r \sim 20 - 21$ mag
object against the Palomar sky.  Norris is equipped with a sensitive,
thinned, backside--illuminated, anti--reflection--coated SITe 2048$^2$
CCD.  Due to the large plate scale (2.55 arcsec/mm) of the Cassegrain
focus of the 5 m telescope, the fibers can be placed within 16 arcsec
of each other, allowing comoving scales as small as $0.1 h^{-1}$ Mpc
to be probed at $z \sim 0.3$.  Smaller scales can be probed by
observing a given field more than once.  Our velocity accuracy, judged
from repeat observations, is $\sim 75$ km s$^{-1}$.

For each galaxy, we estimate a coarse spectral type based on its Gunn
$g-r$ color, where the magnitudes are measured from POSS-II $J$ and
$F$ plates of the fields, and its redshift.  We classify the galaxies
into spectral classes based on the E, Sbc, Scd, and Im spectral energy
distributions compiled by Coleman, Wu, \& Weedman (1980).  The
spectral type is a real number that takes the values of 0 for an
elliptical galaxy, 2 for an Sbc galaxy, 3 for an Scd galaxy, and 4 for
an Im galaxy.  We interpolate between the Coleman et al.\ (1980)
spectral energy distributions to construct the spectral energy
distribution appropriate for the give spectral type.  Finally, we use
the interpolated spectral energy distribution to compute the
$k$--correction necessary to transform between apparent and absolute
magnitude.  (For a more extensive discussion of our spectral
classification and computation of absolute magnitudes, see Small,
Sargent, \& Hamilton [1997b].)

The Norris data suffer from both magnitude and spatial selection
effects.  We assume that the total selection function is separable:
\begin{equation}
S(m,\alpha,\delta) = s_m(m) \times s_{\alpha,\delta}(\alpha,\delta),
\end{equation}
where $s_m(m)$ is the probability that a galaxy with magnitude $m$ is
sufficiently well detected to yield a secure redshift and
$s_{\alpha,\delta}(\alpha,\delta)$, where $\alpha$ and $\delta$ are
celestial coordinates, is the geometrical modulation of $s_m(m)$
(c.f. Yee et al.\ 1996).  The mean value of
$s_{\alpha,\delta}$ is approximately one.  We could also account for
the fraction of objects that are classified as galaxies in the
original catalog from which we select objects but which turn out to be
misclassified stars.  However, since this fraction is small ($\sim
10$\%) and does not vary spatially, we have chosen to ignore it.  The
precise forms of $s_m$ and $s_{\alpha,\delta}$ for the two surveys are
described in the following two subsections.

\subsection{The Corona Borealis Survey}

\begin{figure}
\plotfiddle{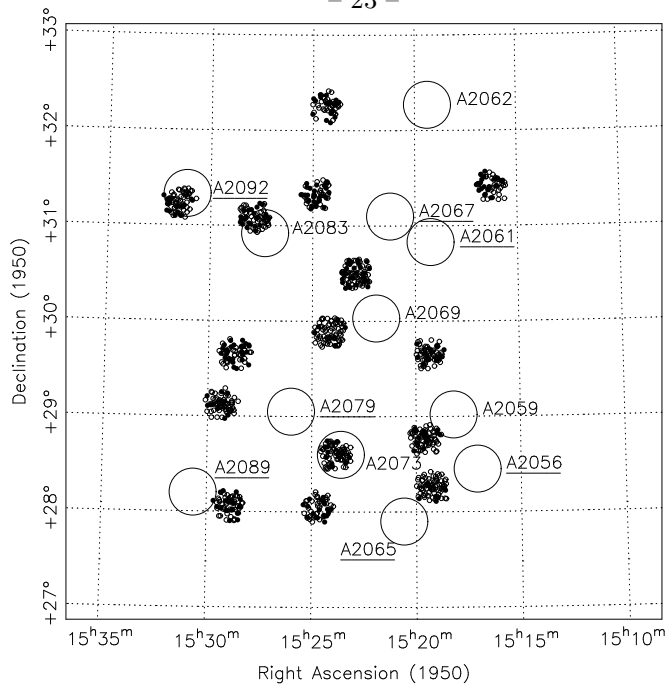}{2.75 in}{-90}{50}{50}{-198}{286}
\caption{
Location on the sky of all 981 $r < 21$ mag galaxies with measured
redshifts in the Corona Borealis survey fields included in this paper.  
Galaxies plotted with filled circles have $z > 0.2$, while galaxies
plotted with unfilled circles have $z < 0.2$.  The large circles
mark the positions of the cataloged Abell clusters in the field.
Clusters whose names are underlined lie within the Corona Borealis
supercluster ($z \approx 0.07$).}
\label{figures:CB_on_sky}
\end{figure}

The survey of the Corona Borealis supercluster has already been
described in the literature (Small, Sargent, \& Hamilton 1997a), and
so we will only briefly review the salient points here.  The core of
the supercluster covers a $6\arcdeg \times 6\arcdeg$ region of the sky
and consists of seven rich Abell clusters.  Since the field of view of
the Norris Spectrograph is 20\arcmin\ in diameter, we planned to
observe 36 fields selected from the POSS-II survey (Reid et al.\ 1991)
and arranged in a rectangular grid with a grid spacing of 1\arcdeg,
with the precise position of a particular field adjusted to maximize
the number of fibers placed on galaxies.  We mainly tried to avoid the
cores of the seven Abell clusters since redshifts for many galaxies in
the cores are available from the literature.  We successfully observed
23 of the program fields and 9 additional fields along the ridge of
galaxies between Abell 2061 and Abell 2067, yielding redshifts for
1491 extragalactic objects.  Our first 17 fields were observed when no
large--format $2048^2$ CCD was available at Palomar, limiting us to
using only one-half of the fibers and leading to geometrical selection
effects for which we are not able to correct.  We have therefore
restricted the analysis presented here to the 981 $r < 21$ mag, $z <
0.5$ galaxies successfully observed with the large--format $2048^2$
CCD available at Palomar since 1994.  Excluding the data from the
first 17 fields observed only reduces the number of galaxies at $0.2 <
z < 0.5$ by 23\%.  The locations on the plane of the sky of all the
galaxies in the Corona Borealis survey are shown in Figure
\ref{figures:CB_on_sky}.  Galaxies with $z > 0.2$ are marked with
filled circles while galaxies with $z \le 0.2$ are marked with
unfilled circles.

\begin{figure}
\plotfiddle{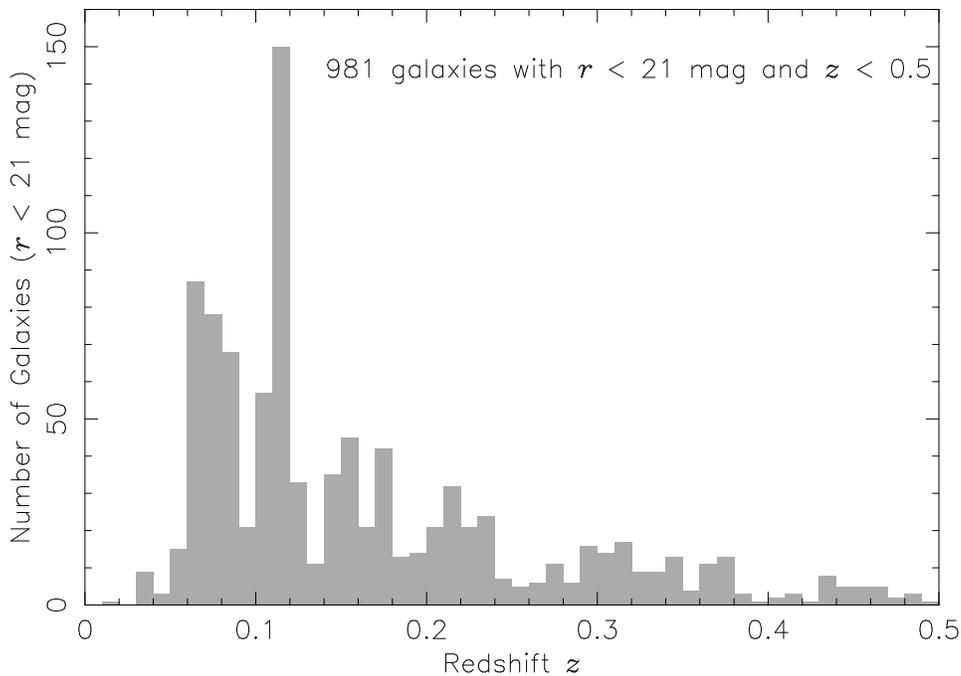}{2.75 in}{-90}{50}{50}{-198}{286}
\caption{
Redshift histogram of $r < 21$ mag galaxies in the Corona Borealis fields
included in this paper.  The two prominent peaks, one at $z \approx 0.07$ and
the other at $z \approx 0.11$, are the Corona Borealis supercluster
and the A2609 supercluster, respectively.  (See Small et al.\ [1998] for a
detailed analysis of the superclusters.)}
\label{figures:CB_z}
\end{figure}

The redshift distribution for all galaxies with $r \le 21$ mag is
shown in Figure \ref{figures:CB_z}.  The prominent features at $z
\approx 0.07$ and $z \approx 0.11$ are the two superclusters in the
field; see Small et al.\ (1998) for a detailed discussion of the
properties of the superclusters.  In our analysis of the field galaxy
luminosity function in the Corona Borealis survey (Small et al.\
1997b), we found that the region from $0 < z \le 0.2$, with the
superclusters {\it excluded}, was overdense by 21\% relative to other
high--Galactic--latitude fields.  While this overdensity is not an
exceptional fluctuation, we have conservatively chosen to avoid the
complications of analyzing such a large overdense region and have
limited our study to galaxies with $0.2 < z < 0.5$.  The galaxy number
density over this redshift range is consistent with the number density
found in the CFRS (Lilly et al.\ 1995b) and CNOCII (Lin et al.\ 1998)
surveys.

\begin{figure}
\plotfiddle{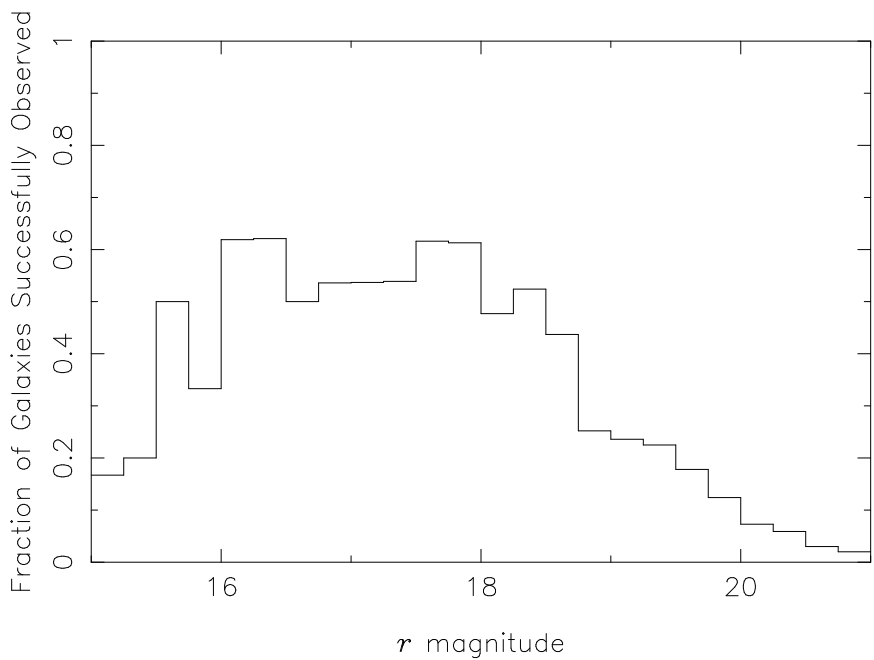}{2.75 in}{-90}{50}{50}{-198}{286}
\caption{
Histogram of the ratio of the number of galaxies with measured
redshifts to the total number of galaxies in the survey fields as a function
of magnitude, $s_m(m)$.}
\label{figures:CB_mag}
\end{figure}

The Corona Borealis survey is not magnitude--limited.  In Figure
\ref{figures:CB_mag}, we plot $s_m(m)$, the ratio of the number of
galaxies with measured redshifts to the total number of galaxies in
the survey fields as a function of magnitude.  For $16 \lesssim r
\lesssim 18.5$, the ratio $s_m(m)$ is nearly constant, but below unity
because of sparse sampling.  It then falls rather steeply to fainter
magnitudes due both to the fiber assignment algorithm and to the
increasing difficulty of measuring redshifts for fainter objects.
Since the light from very bright galaxies can bleed into neighboring
fibers, we have made a modest effort to avoid galaxies brighter than
$r \lesssim 16$ mag, which accounts for the decline in the fraction of
galaxies observed at the brightest magnitudes.  We describe how we
correct for magnitude incompleteness in \S 3.

\begin{figure}
\plotfiddle{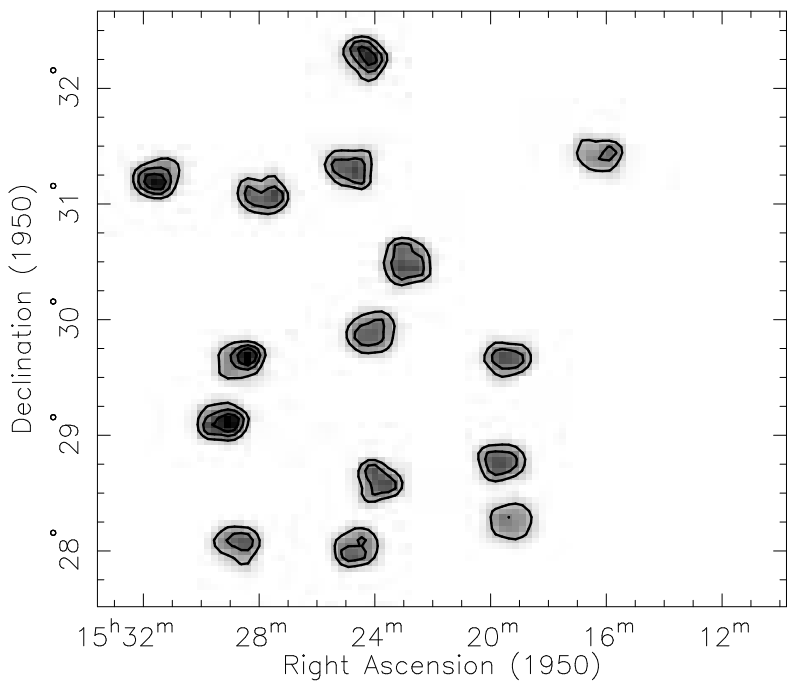}{2.75 in}{-90}{50}{50}{-198}{286}
\caption{
Spatial selection function for the Corona Borealis field.  The gray
scale, ranging from 0. to 2.0, shows the fraction of galaxies
successfully identified in each $3^\prime \times 3^\prime$ area of the
survey field, normalized by the fraction of galaxies identified in the
entire survey.  The contours are drawn at 0.5, 1.0, and 1.5.  For
clarity, we have applied a Gaussian smoothing filter with $\sigma = 1$\arcmin.}
\label{figures:CB_sp_sel}
\end{figure}

Our sampling of the Corona Borealis field on the plane of the sky
varies dramatically.  In particular, much of the $6^\circ \times
6^\circ$ has not been surveyed at all.  We quantify the angular
selection effects by dividing the entire survey area into $103 \times
103$ 9 sq. arcmin cells and, for each cell, computing the ratio of the
number of galaxies with magnitudes $15 \le r \le 21$ with measured
redshifts to the total number of catalog galaxies with magnitudes $15
\le r \le 21$.  The 9 sq.\ arcmin cells are as small as we can make
them while still maintaining a reasonable number of galaxies in each
cell.  The geometrical selection function, $s_{\alpha,\delta}$, is
this ratio normalized by the fraction (16\%) of $15 < r < 21$ magnitude
galaxies in the survey fields for which we have obtained reliable
redshifts.  The mean value of $s_{\alpha,\delta}$ averaged over
the survey fields is 0.99.  We display the map of $s_{\alpha,\delta}$,
Gaussian--smoothed with $\sigma = 1$\arcmin\ for clarity, in Figure
\ref{figures:CB_sp_sel}.  The gray scale ranges linearly from 0. to
2.0, and the contours are drawn at 0.5, 1.0, and 1.5.  The most
prominent feature of the map is the radial dependence of the sampling
within a given Norris field.  The decline in sensitivity at the edges
of the field is due to a combination of effects: mild vignetting,
curvature of the focal plane which makes fibers at the edge slightly
out of focus, and a bias against pairs with large angular separation
introduced by the fiber assignment software (see Small et al.\ 1997a).

\begin{figure}
\plotfiddle{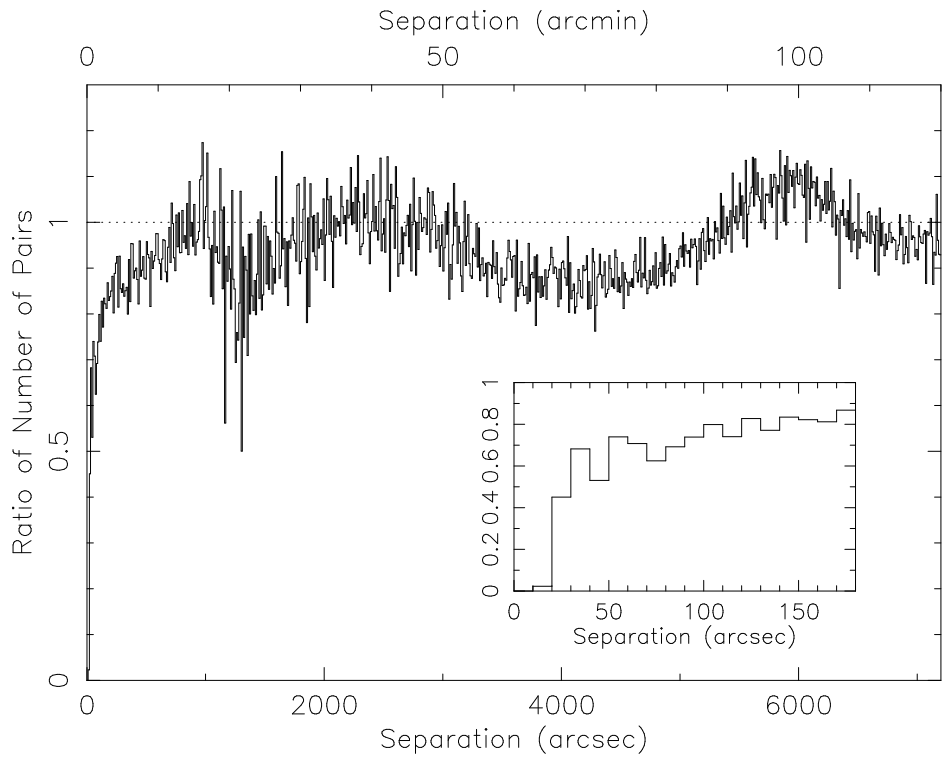}{4.00 in}{-90}{50}{50}{-198}{286}
\caption{
Ratio of number of pairs of galaxies successfully observed to the
number of pairs of galaxies in the catalog selected according to the
combined (magnitude times geometrical) selection function of the
survey.  The inset plot shows an expanded view for separations from 0
to 180 arcsec.}
\label{figures:CB_pairs}
\end{figure}

Since the survey area has been divided into $3^\prime \times 3^\prime$
cells, this procedure does not correct for spatial selection effects
on scales smaller than 3\arcmin.  The fibers cannot be placed within
16\arcsec\ of each other, and so we expect a substantial deficit of
observed pairs on scales smaller than $\sim 50$ arcsec.  This deficit
is illustrated in Figure \ref{figures:CB_pairs} in which we plot, as a
function of pair separation, the ratio of the number of observed pairs
of galaxies to the number of pairs, averaged over 50 realizations, of
galaxies selected from the survey fields according to the total
selection function (i.e., magnitude selection times geometrical
selection).  There is a strong bias against observed pairs with
separations smaller than 100\arcsec.  When computing correlation
functions, we use the ratio shown in Figure \ref{figures:CB_pairs} to
correct the observed pair counts for these missing
small--angular--separation pairs.  (This correction is described in
more detail in \S 3.)  The plot in Figure \ref{figures:CB_pairs} also
reveals a significant deficit of pairs with angular separations
ranging from 13\arcmin\ to 33\arcmin, scales comparable 20\arcmin\
field--of--view of the Norris spectrograph.  Since there are so few
observed pairs with separations on this scale, we cannot accurately
correct for missing pairs on this scale and simply leave a gap in the
correlation function at the physical scale corresponding to this
angular scale ($\sim 5 h^{-1}$ Mpc at $z \sim 0.3$).  We do not
attempt to measure correlations on scales larger than 30\arcmin\ and
therefore ignore the imperfections in our model of the geometrical
selection effects on large scales.

\subsection{Abell 104 Survey}

\begin{figure}
\plotfiddle{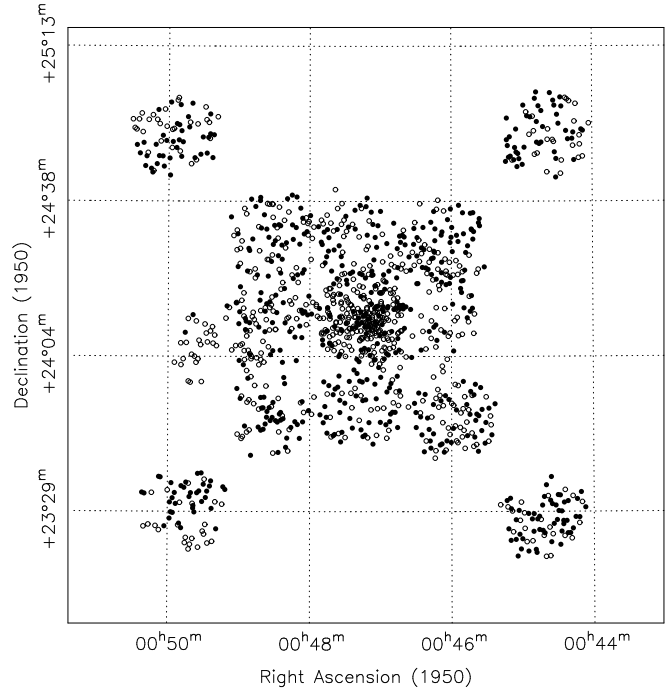}{2.75 in}{-90}{50}{50}{-198}{286}
\caption{
Plot of all galaxies in the Abell 104 survey field with $r < 21$ mag.
Galaxies plotted with filled circles have $z > 0.2$, while galaxies
plotted with unfilled circles have $z < 0.2$.  The Abell 104 cluster
is located at the center of the plot.}
\label{figures:A104_on_sky}
\end{figure}

The data from our Abell 104 survey are similar in most respects to the
data obtained in our Corona Borealis survey.  We have surveyed 1 sq.\
deg.\ centered on Abell 104, plus four outlying fields northeast,
northwest, southeast, and southwest of the center by $\sim 1.5^\circ$.
We have measured spectra for 1330 galaxies in the survey, 207 of which
lie in Abell 104 at $z \approx 0.08$.  We have 558 galaxies with
redshifts $0.2 < z < 0.5$ and magnitudes $15 < r < 21$.  The locations
on the sky of these 558 galaxies (filled circles), along with the
locations of the $r < 21$ mag galaxies with $z < 0.2$ (unfilled
circles), are shown in Figure \ref{figures:A104_on_sky}.  The survey
will be described in detail in Small, Sargent, \& Hamilton (1999).

\begin{figure}
\plotfiddle{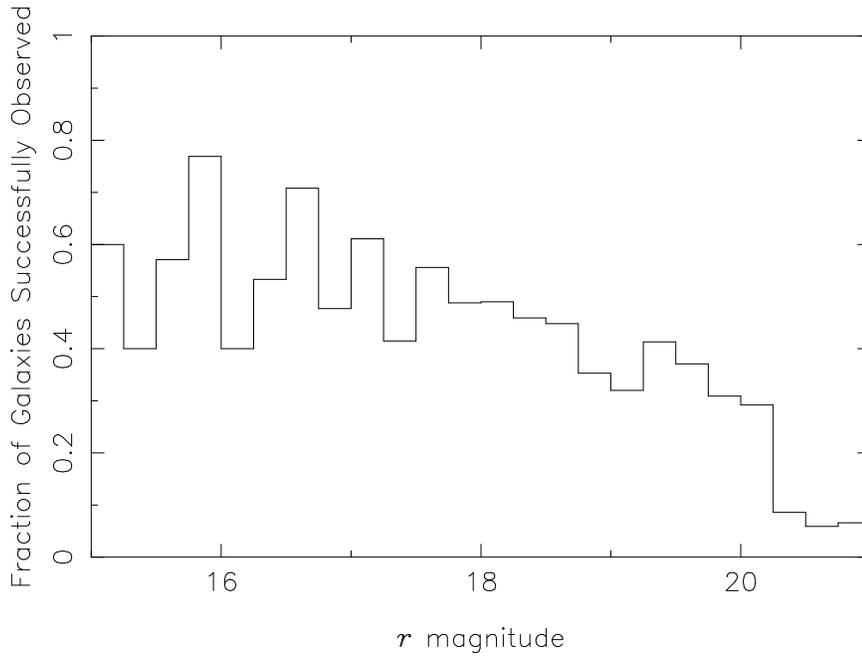}{2.75 in}{-90}{50}{50}{-198}{275}
\caption{
Histogram of the ratio of the number of galaxies with measured
redshifts to the total number of galaxies in the field as a function
of magnitude, $s_m(m)$.}
\label{figures:A104_mag}
\end{figure}

\begin{figure}
\plotfiddle{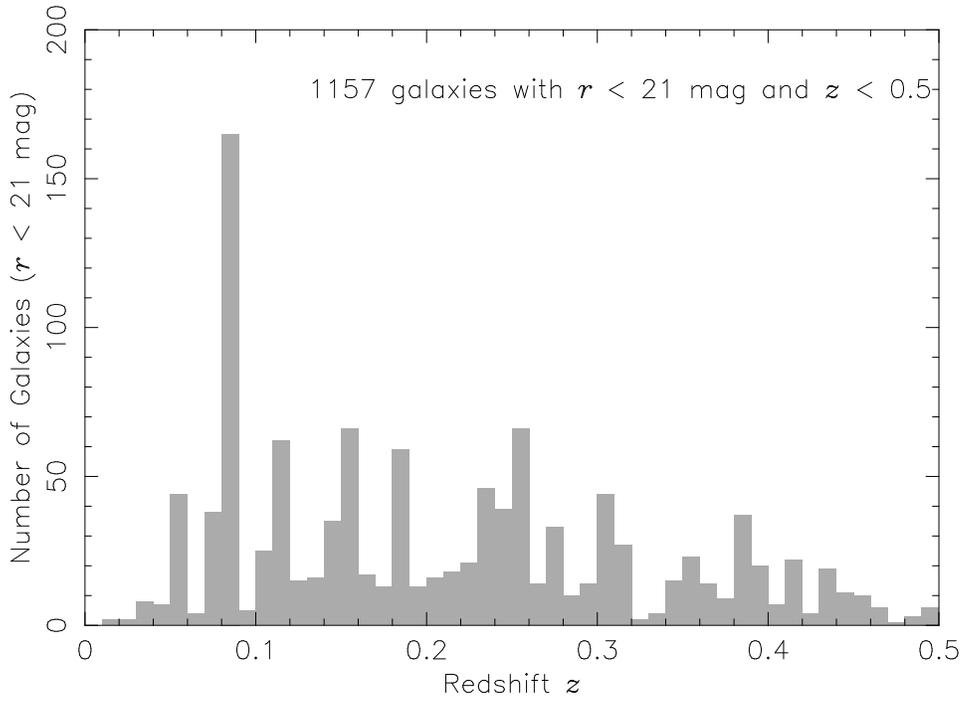}{2.75 in}{-90}{50}{50}{-198}{286}
\caption{
Redshift histogram of galaxies in the Abell 104 survey field with
$r < 21$ mag.  The prominent peak at $z \approx 0.08$ is the
Abell 104 cluster of galaxies.}
\label{figures:A104_z}
\end{figure}

As with the Corona Borealis survey, the A104 survey is not
magnitude--limited.  A plot of $s_m(m)$, the ratio of the number of
galaxies with measured redshifts to the total number of galaxies in
the survey fields as a function of magnitude, is shown in Figure
\ref{figures:A104_mag}.   The A104 field is less sparsely sampled than the
Corona Borealis field.  The magnitude selection function declines
slowly from $r \approx 15$ mag to $r \approx 20$ mag, beyond which the
selection function drops sharply.  The A104 survey is modestly deeper
than the Corona Borealis survey, principally because the experience
gained during the Corona Borealis survey was successfully applied to
the A104 survey.  The increased depth of the A104 survey is borne out
in the redshift distribution plotted in Figure \ref{figures:A104_z}.

\begin{figure}
\plotfiddle{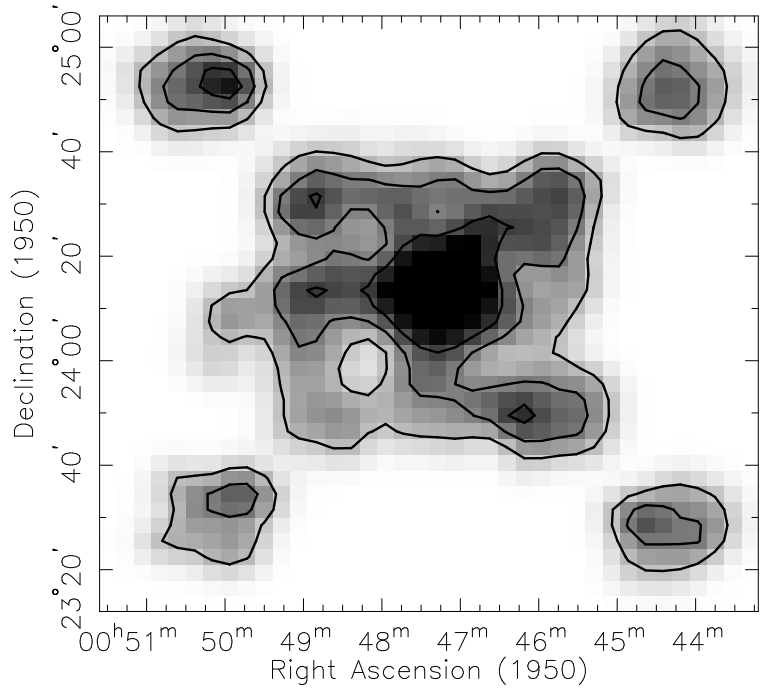}{2.75 in}{-90}{50}{50}{-198}{286}
\caption{
Spatial selection function for the Abell 104 field.  The
gray scale, ranging from 0. to 2.0, shows the fraction of galaxies
successfully identified in each $3^\prime \times 3^\prime$ area of
the survey field.  The contours are drawn at 0.50, 1.0, and 1.5.}
\label{figures:A104_sp_sel}
\end{figure}

\begin{figure}
\plotfiddle{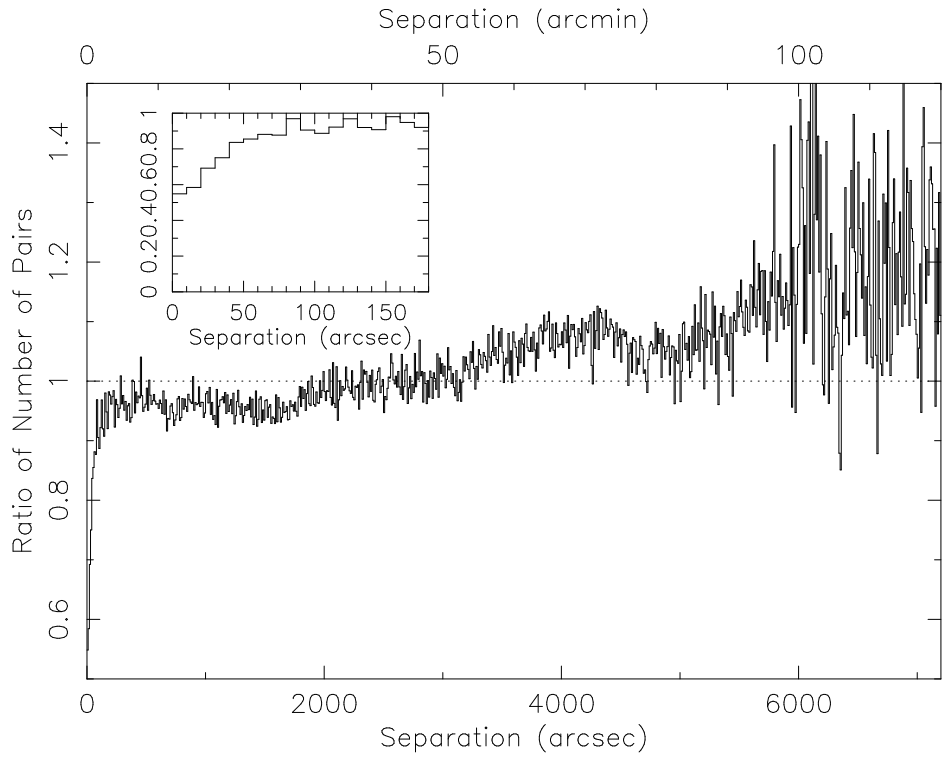}{3.00 in}{-90}{50}{50}{-198}{286}
\caption{
Ratio of number of pairs of galaxies successfully observed to the
number of pairs of galaxies in the catalog selected according to the
combined (magnitude plus spatial) selection function of the survey.
The inset plot shows an expanded view for separations from 0 to 180
arcsec.}
\label{figures:A104_pairs}
\end{figure}

The A104 survey, like the Corona Borealis survey, suffers from uneven
spatial sampling.  We construct a map of the spatial selection
function $s_{\alpha,\delta}$ for galaxies with $15 \le r \le 21$ in
the same manner as for the Corona Borealis survey.  We divide the
entire survey region into $38 \times 38$ square cells, each of which
has an area of 9 sq. arcmin.  For each cell, the angular selection
function is the number of galaxies with measured redshifts with
magnitude $15 \le r \le 21$ divided by the total number of galaxies
with magnitude $15 \le r \le 21$, normalized by the fraction (22\%) of
galaxies in the entire survey for which we have obtained reliable
redshifts.  The mean value of $s_{\alpha,\delta}$ is 0.98.  This map
is shown in Figure \ref{figures:A104_sp_sel}.  The gray scale ranges
linearly from 0. to 2., and the contours are drawn at 0.5, 1.0, and
1.5.  The spatial selection function is higher in the central field
because the central field was observed multiple times.  The multiple
observations of the central field also increase the number of close
pairs, although a significant bias against close pairs remains.  In
Figure \ref{figures:A104_pairs}, we plot, as a function of angular
separation, the ratio of the number of pairs of galaxies successfully
observed to the number of pairs, averaged over 50 realizations, of
galaxies in the parent catalog selected according to the combined
magnitude and spatial selection functions.  Due to the overlapping
Norris fields in the A104 survey, the sampling of pairs as a function
of angular separation is noticeably more uniform than for the Corona
Borealis survey.  In particular, there is no deficit of pairs on
scales comparable to the size of the Norris field--of--view.  Of
course, as for the Corona Borealis survey, the spatial selection
function can only correct for errors on angular scales larger than the
$3^\prime \times 3^\prime$ cell size in which the selection function
was computed.  Thus, there is still a large error on scales smaller
than 180\arcsec, in the sense that close pairs are excluded from the
redshift survey, which is shown in the inset in Figure
\ref{figures:A104_pairs}.  When computing correlation functions, we
use an additional selection function to correct the smallest angular
scales.  This additional function is simply the ratio of pair
separations for separations less than 200\arcsec, multiplied by 1.04
to bring the flat part of the function to a mean value of 1.

\section{Definition and Computation of $\xi(r_p,\pi)$ and $\xi(r)$}

Redshift--space maps of the spatial distribution of galaxies are
distorted by the peculiar motions of galaxies because the measured
redshift of a galaxy is the sum of the Hubble motion of the galaxy
plus the line--of--sight peculiar velocity.  The most prominent
signatures of redshift--space distortions are the ``fingers of God''
seen in redshift surveys of rich clusters of galaxies, in which the
large velocity dispersion of a cluster spreads out the cluster
galaxies along the line--of--sight in redshift space.  On large
scales, coherent infall into overdense regions and outflow from
underdense regions enhance the correlation function.  Since the
velocities on large scales can be simply related to the mean mass
density of the universe $\Omega_0$ with linear theory, an analysis of
redshift space distortions can in principle yield an estimate of
$\Omega_0$ (Sargent \& Turner 1977, Fisher et al.\ 1994, Hamilton
1998).  The distribution of galaxies on the plane of the sky is not,
however, distorted by peculiar velocities.  Thus, correlation
functions, which one assumes are isotropic in real space when averaged
over sufficiently large volumes, are anisotropic in redshift space.
It is, therefore, useful to compute correlation functions as functions
of separations along the line--of--sight ($r_\pi$) and perpendicular
to the line--of--sight ($r_p$).

The two-point correlation function $\xi(r_p,r_\pi)$ is defined
implicitly by the following equation for the joint probability $\delta
P$ of finding a galaxy in each of two volume elements $dV_1,dV_2$
separated by $r_p$ and $r_\pi$,
\begin{equation}
\delta P = \bar n^2[1+\xi(r_p,r_\pi)]dV_1dV_2,
\end{equation}
where $\bar n$ is the mean galaxy density.  In order to compute
$\xi(r_p,r_\pi)$, we construct a catalog of randomly distributed points
with the same selection function as the real data.  We estimate
$\xi(r_p,r_\pi)$ using the estimator derived, tested, and recommended
by Landy \& Szalay (1993):
\begin{equation}
\xi(r_p,r_\pi) = { DD(r_p,r_\pi) - 2DR(r_p,r_\pi) + RR(r_p,r_\pi) \over {RR(r_p,r_\pi)}},
\end{equation}
where DD$(r_p,r_\pi)$, RR$(r_p,r_\pi)$, and DR$(r_p,r_\pi)$ are the
number of data--data, random--random, and data--random pairs,
respectively, with separations $r_p$ and $r_\pi$.  There are three
important virtues of Landy \& Szalay's estimator: it is affected only
in second order by density fluctuations on the scale of the survey; it
does not require an independent measurement of the mean density of the
survey; and its errors are very nearly Poissonian for an unclustered
population.  

The data--data pair counts are corrected for missing pairs on small
scales using the curves shown in Figures \ref{figures:CB_pairs} and
\ref{figures:A104_pairs}.  As expected and demonstrated below, the
correction works very well for the {\it angular} correlation function.
In applying this correction to the spatial correlation function, we
simply assume, since the bias against close separation pairs is
primarily due to limits on how closely fibers can be placed in the
focal plane of the spectrograph, that the redshift distribution of
unobserved close pairs is identical to that of observed close pairs.

In order to compute the real space correlation function $\xi(r)$, we
follow Davis \& Peebles (1983), and many subsequent workers, by
projecting $\xi(r_p,r_\pi)$ onto the $r_p$ axis.  The projection
$w_p(r_p)$ depends only on the real space correlation function:
\begin{equation}
w_p(r_p)  =  2\int_0^\infty \xi(r_p,r_\pi) dr_\pi
          =  2\int_0^\infty \xi\lbrack(r_p^2 + y^2)^{1/2}\rbrack dy,
\end{equation}
where $y$ is the line--of--sight separation in real space.  The
integrand in the second expression for $w_p(r_p)$ is the correlation
function in real space.  If we assume that $\xi(r) =
(r/r_0)^{-\gamma}$, where $r_0$ is the correlation length and $\gamma$
is the power--law index, the integral for $w_p(r_p)$ can be evaluated
analytically to give:
\begin{equation}
w_p(r_p) = r_p{r_0 \overwithdelims () r_p}^\gamma
{\Gamma {1 \overwithdelims () 2} \Gamma {\gamma - 1 \overwithdelims () 2}
\over {\Gamma{\gamma \overwithdelims () 2}}},
\end{equation}
where $\Gamma$ is the standard gamma function.  By fitting a power law
to $w_p(r_p)$, we can determine the correlation length $r_0$ and power law
index $\gamma$ of the real space correlation function.

We calculate the error in $\xi(r_p,r_\pi)$ using the standard
technique of bootstrap resampling the data (Ling, Frenk, \& Barrow
1986).  We perform 50 bootstrap resamples with replacement and take
the error bars on $\xi(r_p,r_\pi)$ to be the standard deviation of the
bootstrap estimates.  The bootstrap error bars are typically 75--100\%
larger than error bars derived from the standard Poisson estimate,
$\sigma(\xi) = (1+\xi)/\sqrt{DD}$.

The construction of the random catalog is complicated by our magnitude
selection effects and our uneven spatial sampling.  As noted above,
our sample of galaxies is not magnitude-limited.  In addition, it is
important that the color distribution of the galaxies in the random
catalog matches the observed color distribution.  We have, therefore,
selected the redshifts of galaxies in the random catalog not from the
probability distribution $P(M \vert z)$ that an object at redshift $z$
has an absolute magnitude $M$, which would be appropriate for a
magnitude--limited sample, but rather from the distribution $P(z \vert
m, type)$ that a galaxy with an apparent magnitude $m$ and spectral
type $type$ (rounded to E, Sbc, Scd, or Im; see \S 2) has a redshift $z$
(suggested by D. Hogg, personal communication):
\begin{equation}
P(z \vert m, type) = {\phi \lbrack M(z,m,type) \rbrack {dV \over {dz}} \over
{\int_0^\infty \phi \lbrack M(z^\prime,m,type) \rbrack {dV \over {dz^\prime}} 
dz^\prime}}.
\label{equations:p_z_m}
\end{equation}
Here, $\phi(M)$ is the luminosity function, $M(z,m,type)$ is the
absolute magnitude of a galaxy of spectral type $type$ such that it
would have apparent magnitude $m$ at redshift $z$, and $dV/dz$ is the
comoving relativistic volume element.  For each galaxy in the survey,
we generate 100 galaxies in the random catalog with the same apparent
magnitude and spectral type as the given galaxy, with redshifts drawn
according to Equation \ref{equations:p_z_m}, and with celestial
coordinates distributed according to the spatial sampling maps
presented in \S 2.  Thus, the distribution of apparent magnitudes and
spectral types of the galaxies in the random catalog is identical to
that of the real survey data, but the locations are random.  Note that
when analyzing a subset of the survey data selected by spectral type
or intrinsic luminosity, it is trivial with this method to ensure that
the random catalog is generated with exactly the same selection
function.

We use a Schechter (1976) function to describe the luminosity function
of our survey.  For $0.2 < z < 0.5$, we use Schechter parameters
$\alpha = -0.85$ and $M^\ast = -19.45 + 5\log h$ ($q_0 = 0.5$) in the
$B_{AB}$--band, consistent with the results from the CNOCII survey
(Lin et al.\ 1998) and our own survey (Small et al.\ 1997b).
Luminosity functions constructed from the combined A104 and Corona
Borealis surveys agree very well with a Schechter function with these
parameters.  The normalization of the luminosity function drops out of
Equation \ref{equations:p_z_m}.

\begin{figure}
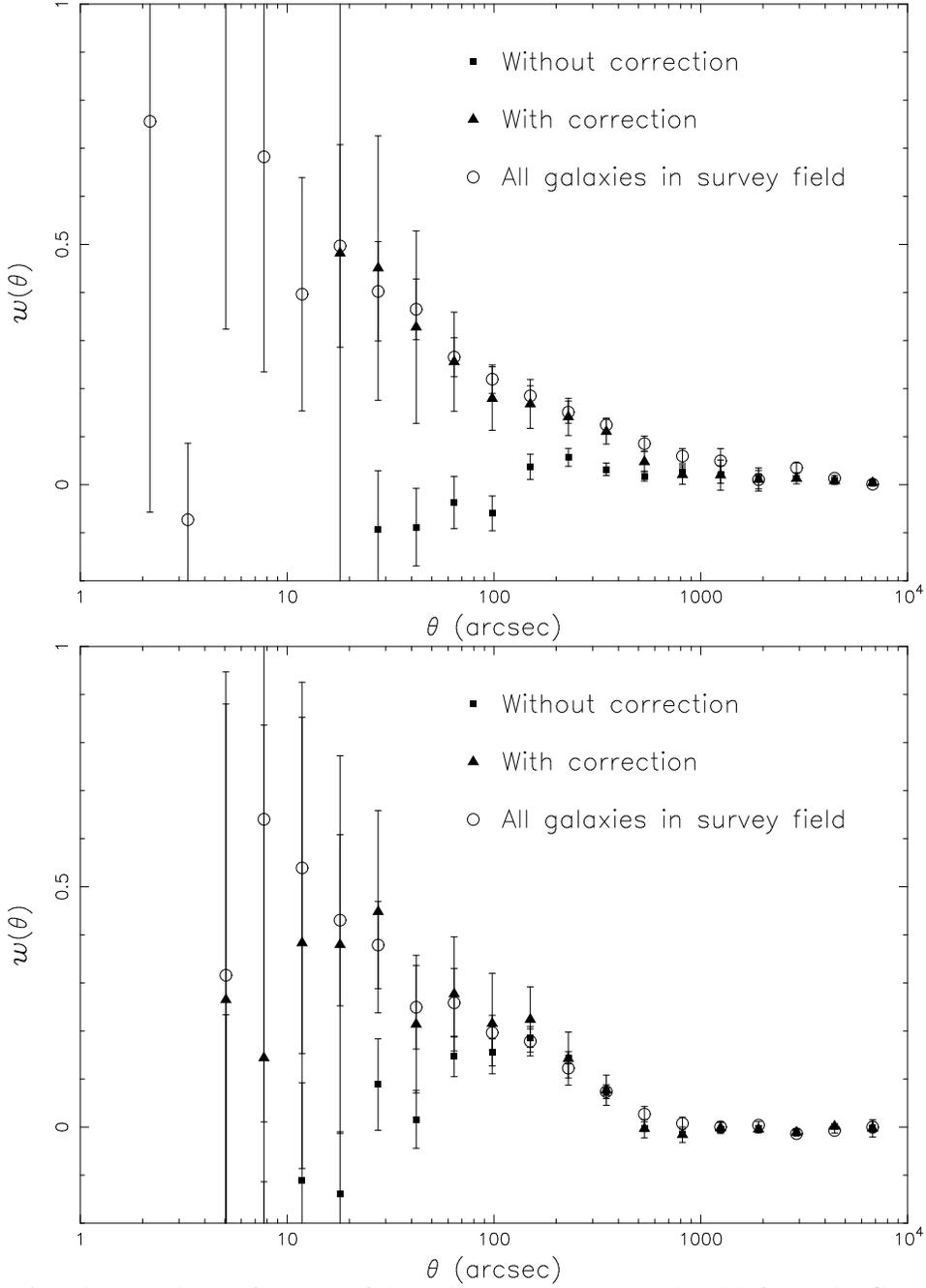

\plotfiddle{figure11a.ps}{3.00 in}{-90}{50}{50}{-198}{296}
\plotfiddle{figure11b.ps}{3.00 in}{-90}{50}{50}{-198}{276}
\caption{
Angular correlation function of the galaxies with measured redshifts
in the Corona Borealis and Abell 104 survey fields (top and bottom
panels, respectively), both with and without correction for missing
pairs on small scales ($\le 600$\arcsec\ for Corona Borealis and $\le
200$\arcsec\ for Abell 104; filled triangles and squares,
respectively), compared with that of all the galaxies in the survey
field (unfilled circles).  With the correction for missing pairs, the
angular correlation functions of the galaxies with measured redshifts
agree well with the angular correlation functions of all the galaxies
in the surveys.}
\label{figures:w_theta}
\end{figure}

Since our method for generating the redshifts of the galaxies in the
random catalog is novel and our spatial selection effects are quite
dramatic, we have conducted two tests to verify that our techniques
are working correctly.  First, to test our corrections for our spatial
selection effects, we have compared the two--point angular correlation
function, $\omega(\theta)$, of galaxies selected from the parent
photometric catalog according to the magnitude selection function
(referred to as the ``photometric sample'' below) with the angular
correlation function of the galaxies with measured redshifts
(c.f., Shepherd et al.\ 1997).  We have estimated $\omega(\theta)$
using the Landy--Szalay estimator.  In order to estimate the errors
for the angular correlation function of the photometric sample, we
have computed the angular correlation function for 50 samples selected
from the photometric catalog and averaged the results.  The angular
correlation functions for the Corona Borealis and Abell 104
photometric and redshift samples are plotted in Figure
\ref{figures:w_theta}.  The correlation functions of the redshift
samples, with bootstrap error bars, are plotted both with and without
corrections for missing pairs on small scales ($< 600$\arcsec\ for the
Corona Borealis survey and $< 200$\arcsec\ for the Abell 104 survey).
Without this correction, the correlation functions fall significantly
below the correlation functions of the photometric samples on small
scales.  With the correction, the agreement between the two
correlation functions for the photometric and redshift samples over
all scales is excellent.  Assuming that the redshift distribution of
the missing pairs on small scales is similar to that of the
successfully observed pairs, then our success at correcting for
spatial selection effects in the angular correlation function should
carry over to the spatial correlation function.

We have also tested our method for generating the redshift
distribution by computing the spatial correlation function with a
random catalog generated by standard methods.  We place galaxies in
the random catalog with uniform comoving density and choose the
magnitudes of the galaxies from a Schechter luminosity function with
the same Schechter parameters $\alpha$ and $M^\ast$ as used above.
The galaxies in the random catalog are then rejected according to the
magnitude and spatial selection effects of the real data.  The
correlation functions computed with this random catalog agree well
with correlation functions computed with our method outlined above,
but we prefer our method because it more naturally allows us to
generate a sample with the correct color distribution as well as
redshift distribution.

\section{$\xi$ for the Abell 104 and
Corona Borealis Fields}

\begin{figure}
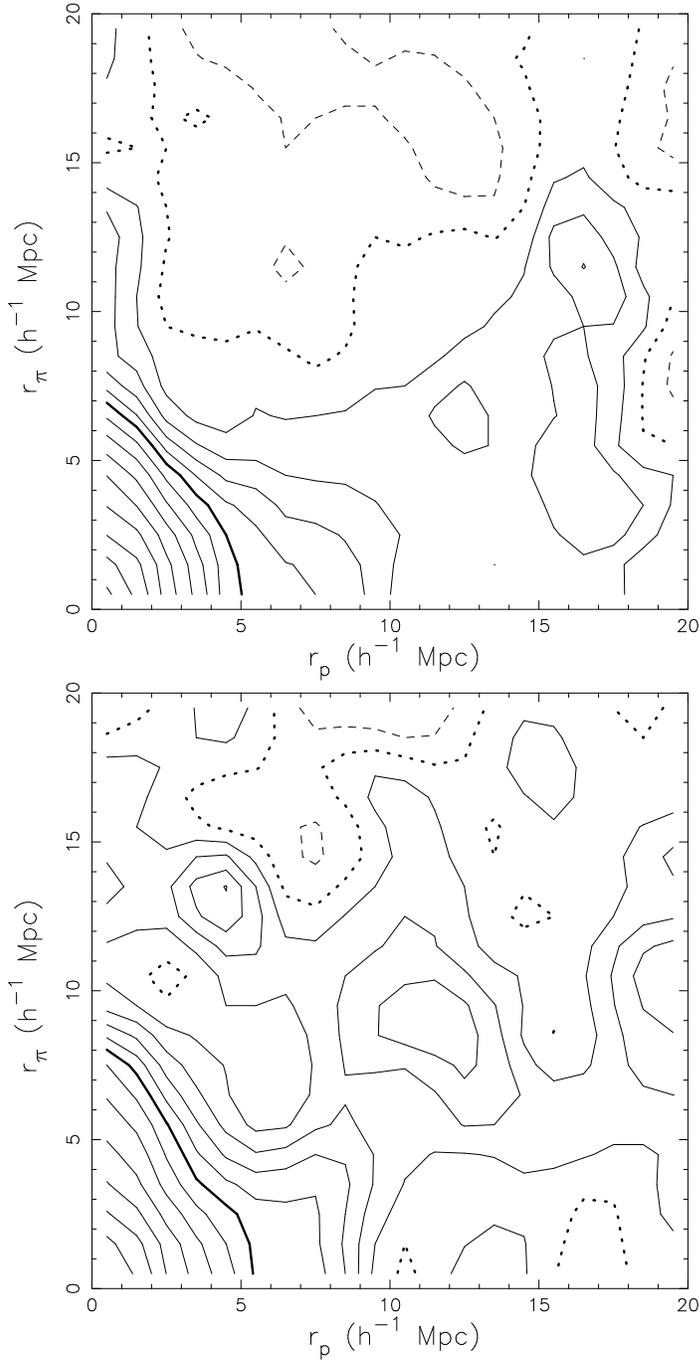

\plotfiddle{figure12a.ps}{3.00 in}{-90}{50}{50}{-198}{312}
\plotfiddle{figure12b.ps}{3.00 in}{-90}{50}{50}{-198}{286}
\caption{
Smoothed contour plots of $\xi(r_p,r_\pi)$ for Abell 104 (top panel)
and Corona Borealis (bottom panel) galaxies with $0.2 \le z \le 0.5$.
The contour levels are described in the text.  The elongation of the
contours at small $r_p$ is due to the velocity dispersion of bound
paris.  The flattening of the contours for $5h^{-1}\rm{Mpc} \lesssim
{r_p} \lesssim 10h^{-1}\rm{Mpc}$ is due to coherent motions.  For the
Abell 104 survey, $\xi(r_p,r_\pi)$ is dominated by only a few
structures and is very unlikely to be representative of the universe
as a whole for $r_p \gtrsim 10h^{-1}$ Mpc.}
\label{figures:contour}
\end{figure}

In Figure \ref{figures:contour}, we show $\xi(r_p,r_\pi)$ for galaxies
with $0.2 \le z \le 0.5$ in the Abell 104 (top panel) and Corona
Borealis (bottom panel) fields.  The thick dark line denotes
$\xi(r_p,r_\pi) = 1$.  Contours above $\xi(r_p,r_\pi) = 1$ are spaced
in logarithmic intervals of 0.1 dex, while contours below
$\xi(r_p,r_\pi)$ are spaced in linear intervals of 0.2 with
$\xi(r_p,r_\pi) = 0$ marked with a heavy dashed line.
$\xi(r_p,r_\pi)$ has been computed here in linear bins of $1 h^{-1}$
comoving Mpc.  For clarity of presentation, to emphasize the most
important features of the data, and to reduce binning noise, we have
smoothed (twice in the case of the Corona Borealis data) the
correlation functions shown here with a $3 \times 3$ filter:
\begin{equation}
\left(
\begin{array}{ccc}
0.75 & 1.00 & 0.75 \\
1.00 & 2.00 & 1.00 \\
0.75 & 1.00 & 0.75
\end{array}
\right)
\end{equation}
Although the signal--to--noise ratio is significantly higher for the
A104 field than for the Corona Borealis field, the same principal
features are apparent in both correlation functions.  For small $r_p$,
there is dramatic elongation of the contours along the $r_\pi$ axis
due to the velocity dispersion of bound pairs.  At larger $r_p$, there
is a substantial compression of the contours due to coherent motions.
In the Abell 104 field, only a few structures contribute to
$\xi(r_p,r_\pi)$ for $r_p \gtrsim 10h^{-1}$ Mpc, certainly making the
results on these scales unreliable.  For example, removing all
galaxies from the Abell 104 sample with $0.24 < z < 0.27$, and thus
removing the prominent shell structure in the galaxy distribution (see
Figure \ref{figures:A104_pie} below), reduces $\xi(r_p,r_\pi)$ at $r_p
\approx 15h^{-1}$ Mpc to zero within the errors.  At smaller projected
separations, we do not see any features in $\xi(r_p,r_\pi)$ in either
field that can be associated with individual structures in the galaxy
distribution.  This is not surprising since the depth of the surveys
along the line of sight ($\Delta z = 0.3$, corresponding to a comoving
depth of $657,\ 578 h^{-1}$ Mpc for $q_0 = 0.1,\ 0.5$) is
substantially greater than the largest structures in the galaxy
distribution, which have sizes of typically $50 - 60 h^{-1}$ Mpc (see
\S 6).

\begin{figure}
\plotfiddle{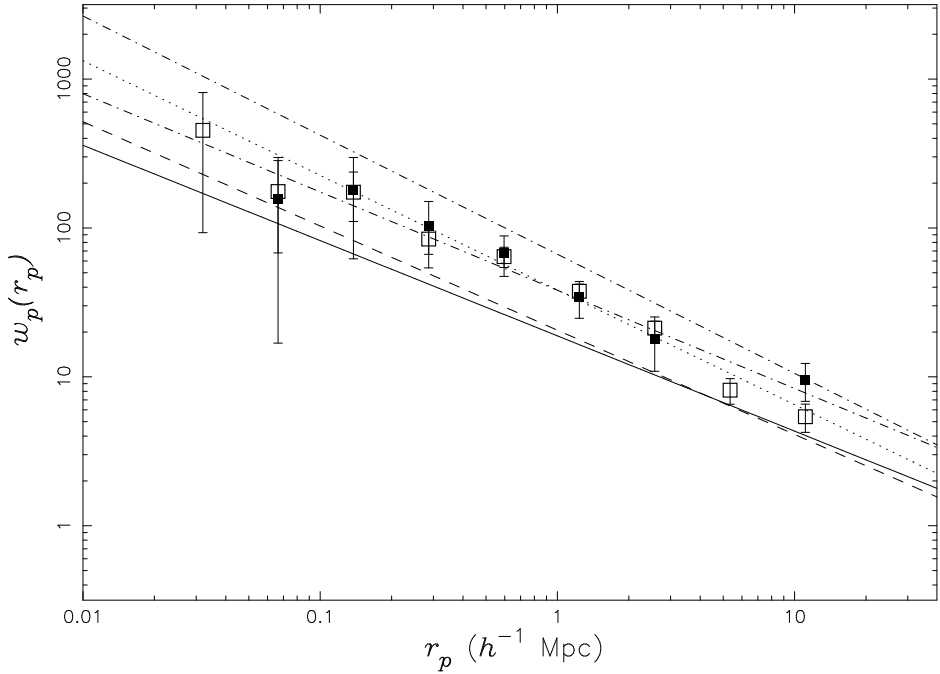}{3.00 in}{-90}{50}{50}{-198}{300}
\caption{
$w_p(r_p)$ for $0.2 \le z \le 0.5$ for the Abell 104 survey (unfilled
squares) and the Corona Borealis survey (filled squares).  Also shown
are power--law fits to the correlation functions computed for the CFRS
(solid line) and CNOC1 (dashed line) surveys, local
optically--selected surveys (upper dash--dotted line), the 1.2 Jy
$IRAS$ Galaxy Redshift survey (lower dash--dotted line), and the best
fit to the Abell 104 survey data (dotted line).}
\label{figures:wp}
\end{figure}

\begin{figure}
\plotfiddle{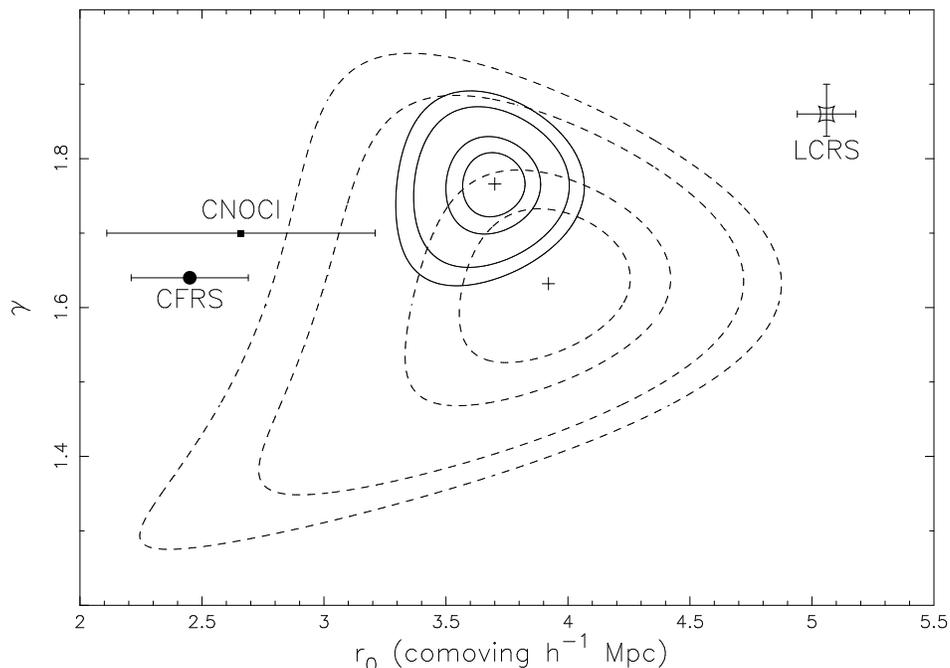}{3.00 in}{-90}{50}{50}{-198}{276}
\caption{
Error contours of $(r_0, \gamma)$ for power--law fits to the projected
two--point correlation functions of galaxies with $0.2 < z < 0.5$ in
the Abell 104 (solid contours) and Corona Borealis (dashed contours)
fields.  The contour levels correspond to 1$\sigma$ on the parameters
taken individually, 1$\sigma$ on the parameters taken jointly,
2$\sigma$ on the parameters taken jointly, and 3$\sigma$ on the
parameters taken individually.  Also plotted are the values of $r_0$
and $\gamma$ from the CFRS survey (with $\gamma$ held fixed at 1.64), the
CNOCI survey (with $\gamma$ held fixed at 1.70), and the Las Campanas
Redshift survey.}
\label{figures:r0_gamma}
\end{figure}

The projected correlation function, $w_p(r_p)$, is the integral of the
correlation functions (see Equation 4) shown in Figures
\ref{figures:contour} along the $r_\pi$ axis.  For the computation of
$w_p(r_p)$, we recompute $\xi(r_p,r_\pi)$ using logarithmic bins in
$r_p$.  In Figure \ref{figures:wp}, we plot $w_p(r_p)$ for galaxies
with $0.2 < z < 0.5$ ($z_{med} = 0.30$) in the A104 field (unfilled
squares) and the Corona Borealis field (filled squares).  We have
integrated $\xi(r_p,r_\pi)$ along the $r_\pi$ axis out to $r_{\pi,max}
= 15 h^{-1}$ Mpc.  The results are insensitive to $r_{\pi,max}$ within
the errors.  The correlation functions for the two fields agree very
well, which strongly suggests that we are obtaining a fair estimate of
the $0.2 < z < 0.5$ correlation function.  Both correlation functions
are well fit by a power law correlation function in real space.  We
obtain $r_0 = 3.70 \pm 0.13 h^{-1}$ comoving Mpc and $\gamma = 1.77
\pm 0.05$ for the Abell 104 field, and we obtain $r_0 = 3.92 \pm 0.35
h^{-1}$ comoving Mpc and $\gamma = 1.63 \pm 0.10$ for the Corona
Borealis field.  Note that there is no data point plotted at $r_p
\approx 5 h^{-1}$ Mpc for the projected correlation function of the
Corona Borealis field since this is the physical scale that
corresponds to the field of view of the spectrograph at $z \sim 0.3$,
and we are not able to construct a reliable geometrical selection
function on this scale (\S 2.1).  This bias has negligible effect on
the surrounding bins.  The best fit to the Abell 104 survey data is
plotted with a dotted line.  Error contours for ($r_0, \gamma$) are
plotted in Figure \ref{figures:r0_gamma}.  Our results are summarized
in Table 1, in which, for each sample analyzed, we list the redshift
range, the number of galaxies in the sample, the median redshift,
$r_0$ for $q_0 = 0.1$ and $q_0 = 0.5$, and the power--law index
$\gamma$ (which varies negligibly with changing $q_0$).

As we discussed in the Introduction and discuss in more detail below,
different populations of galaxies cluster differently.  It is
especially important to bear this fact in mind when comparing the
clustering of populations at different redshifts since one may not, in
fact, be observing the same population at all redshifts.  We do not
have detailed morphological information for the galaxies in our
sample, but we do know that there are 2-3 times more star--forming
galaxies at $0.2 < z < 0.5$ than at $z < 0.2$ (Small et al.\ 1997b).
Our sample at $0.2 < z < 0.5$ {\it may}, therefore, have a mix of
galaxy types that is closer to the mix in a local sample selected by
infrared rather than optical luminosity.

We have also plotted in Figure \ref{figures:wp} the results from the
CFRS (Le~ F\`evre et al.\ 1996, solid line) and from the CNOCI
(Shepherd et al.\ 1997, dashed line) surveys for the correlation
function of galaxies with $0.2 < z < 0.5$.  It is immediately apparent
that our results imply a substantially larger correlation length than
obtained in the two earlier surveys.  Nevertheless, our results still
indicate significant evolution relative to the comoving correlation
length measured in local, optically--selected surveys, a
representative model ($r_0 = 5.0 h^{-1}$ Mpc and $\gamma = 1.80$) of
which is plotted with the upper dash--dotted line in Figure
\ref{figures:wp}.  The correlation function measured by Fisher et al.\
(1994) for a sample of $IRAS$--selected galaxies is plotted with the
lower dash--dotted line.  The correlation length that we measure is
$0.8 h^{-1}$ Mpc shorter, but only at the $1 \sigma$ significance
level, than the value inferred by Postman et al.\ (1998) for $z = 0.5$
from their wide--field imaging survey.  (See Figure
\ref{figures:r0_evol} for a summary plot.)

\begin{figure}
\plotfiddle{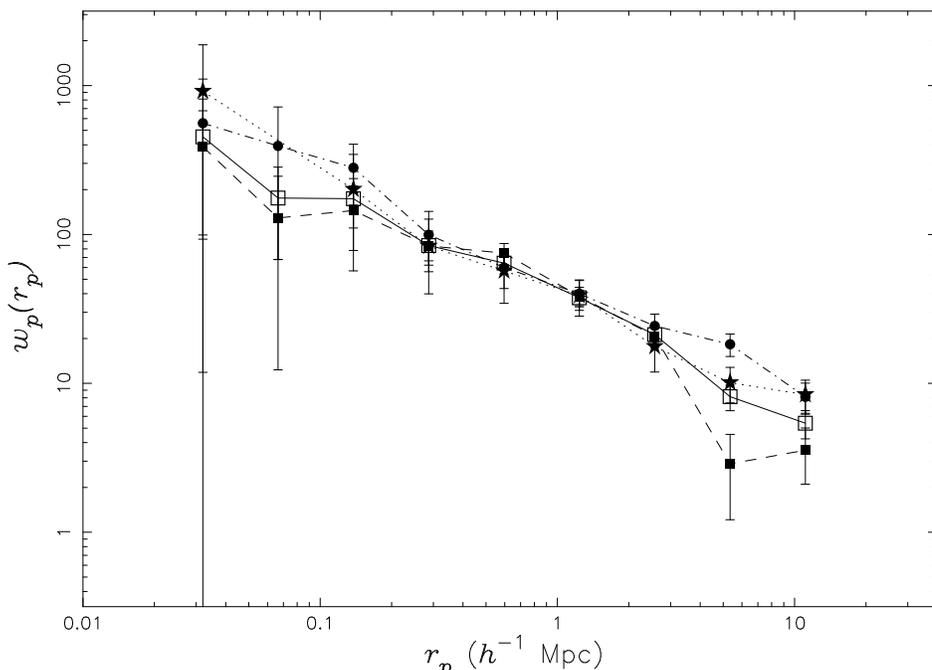}{3.00 in}{-90}{50}{50}{-198}{276}
\caption{
The projected correlation function of galaxy samples from the Abell
104 survey with $0.2 < z < 0.3$ (filled squares), $0.3 < z < 0.5$
(filled circles), $0.32 < z < 0.5$ (filled stars), and $0.2 < z <
0.5$ (unfilled squares)}
\label{figures:wp_z}
\end{figure}

\begin{figure}
\plotfiddle{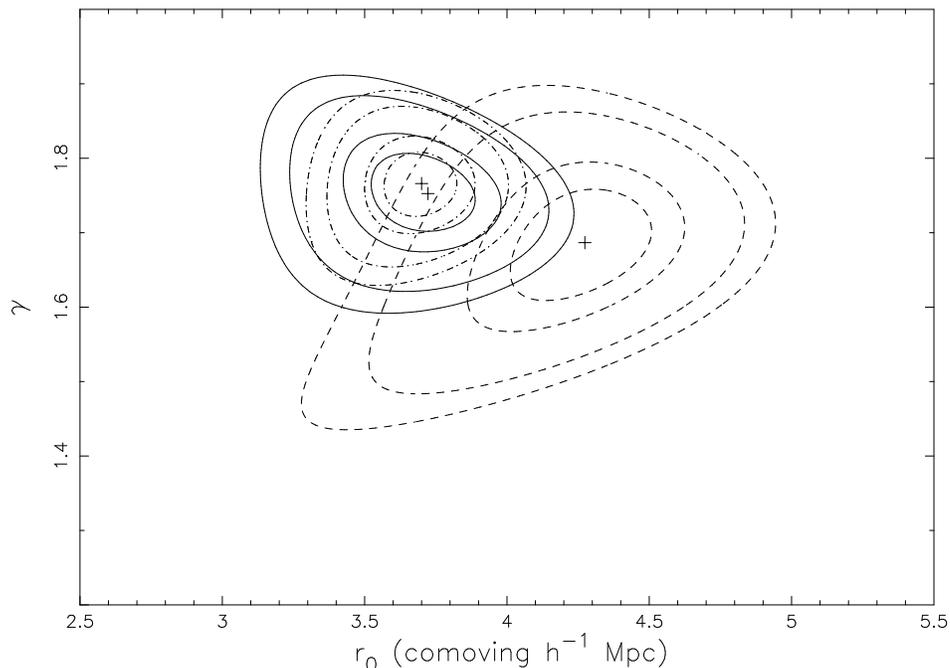}{3.00 in}{-90}{50}{50}{-198}{276}
\caption{
Error contours of $(r_0, \gamma)$ for power--law fits to the projected
correlation functions for samples selected from different redshift
intervals ($0.2 < z \le 0.3$, solid contours; $0.3 < z \le 0.5$,
dashed contours; $0.2 < z < 0.5$, dot--dashed contours) from the Abell
104 survey.  The contour levels are the same as in Figure
\ref{figures:r0_gamma}.  Within the errors, there is no evidence
within our survey for variation of either $r_0$ or $\gamma$ with
redshift.}
\label{figures:r0_gamma_evol} 
\end{figure}

By dividing the A104 survey into smaller redshift intervals, we have
looked for evolution of clustering within our own sample from a median
redshift of $z_{\rm med} = 0.25$ to $z_{\rm med} = 0.39$.  (The Corona
Borealis survey does not contain enough galaxies to be divided into
smaller redshift intervals.)  In Figure \ref{figures:wp_z}, we plot
the projected correlation function of galaxies with redshifts $0.2 < z
\le 0.3$ (filled squares, $z_{\rm med} = 0.25$), $0.3 < z \le 0.5$
(filled circles, $z_{\rm med} = 0.38$), $0.32 < z \le 0.5$ (filled
stars, $z_{\rm med} = 0.39$), and, for reference, $0.2 < z \le 0.5$
(unfilled squares, $z_{\rm med} = 0.30$).  We have included two higher
redshift intervals, one starting at $z = 0.30$ and the other at $z =
0.32$, in order to illustrate the effect one large structure, the
prominent clump of galaxies at $z \approx 0.31$, can have on the
measured correlation function.  There is no discernible effect at
scales smaller than $1 h^{-1}$ Mpc, but, at $r_p = 5.3 h^{-1}$ Mpc,
the projected correlation function of the interval including the large
clump is $1.8 \pm 0.6$ times larger than that of the interval
excluding the clump.  The computed power--law parameters are not
significantly affected by the large structure, with the correlation
lengths and power--law slopes differing at the $1.2\sigma$ and
$0.3\sigma$ levels, respectively.  Within the errors, we see no
evidence for evolution of the projected correlation function between
$z_{\rm med} = 0.25$ and $z_{\rm med} = 0.38$.  The error contours of
power--law fits to the projected correlation functions for galaxies
selected from the intervals $0.2 < z \le 0.3$, $0.3 < z \le 0.5$, and
$0.2 < z \le 0.5$ are shown in Figure \ref{figures:r0_gamma_evol} and
confirm that there is no statistically significant variation with
redshift in the comoving correlation length and power--law slope
apparent in our data.

The difference between the correlation function measured here and that
measured in the CFRS survey is likely due to the differences in the two galaxy
samples and to the small area of the CFRS survey.  The CFRS sample at
$0.2 < z < 0.5$ is mainly sub--$L^\ast$ galaxies, whereas our sample
is concentrated near $L^\ast$.  In the local universe, lower
luminosity galaxies cluster less strongly than higher luminosity
galaxies (Loveday et al.\ 1995), and the difference in clustering
strengths between our sample and the CFRS sample strongly suggests
that this trend continues to intermediate redshifts.  Indeed, Le
F\`evre et al.\ (1996) cautioned that the clustering strength of a
brighter sample of galaxies at $0.2 \le z \le 0.5$ might be
substantially higher than for their sub--$L^\ast$ sample.  The small
area of the CFRS survey, only 114 sq. arcmin, probably also contributes
to reducing the CFRS correlation length.  It is more difficult to
understand the disagreement with the CNOCI survey results, however, as their
sample has a similar range of intrinsic luminosities.  It is possible
that their analysis suffers from too small a field and from the
treatment of the rich galaxy cluster within their field.  Preliminary
results from the CNOCII survey (Carlberg et al.\ 1999) are in better
agreement with our results.  At $z = 0.28$, Carlberg et al.\ (1999)
find $r_0 = 4.86 \pm 0.32 h^{-1}$ Mpc for luminous ($\gtrsim L^\ast$)
galaxies ($q_0 = 0.1$). 

It is well--known that the clustering properties of galaxies in the
local universe depend on the type of galaxies in question.  In
particular, red, early--type galaxies cluster considerably more
strongly than blue, late--type galaxies, and intrinsically bright
galaxies cluster more strongly than intrinsically faint galaxies
(Loveday et al.\ 1995, Guzzo et al. 1997).  We can use our data to
explore whether these trends continue at higher redshift.

Since we do not have accurate morphological information for most of
the galaxies in our survey and our magnitude errors are relatively
large, we have approximated the division of samples by color and
morphological type by dividing our sample by the rest--frame
equivalent width of [\ion{O}{2}]$\lambda3727$.  As discussed in detail
by Kennicutt (1992) and used extensively in galaxy evolution research,
the rest--frame equivalent width of [\ion{O}{2}] is an approximate
measure of the current star formation rate in a galaxy.  We divide our
Abell 104 survey sample at $W_0[$\ion{O}{2}] $= 10$\AA, which
corresponds roughly to dividing the sample at a morphological type of
Sbc (see Figure 11 of Kennicutt [1992]).  The Corona Borealis sample
is, again, too small to be usefully divided into subsamples.

\begin{figure}
\plotfiddle{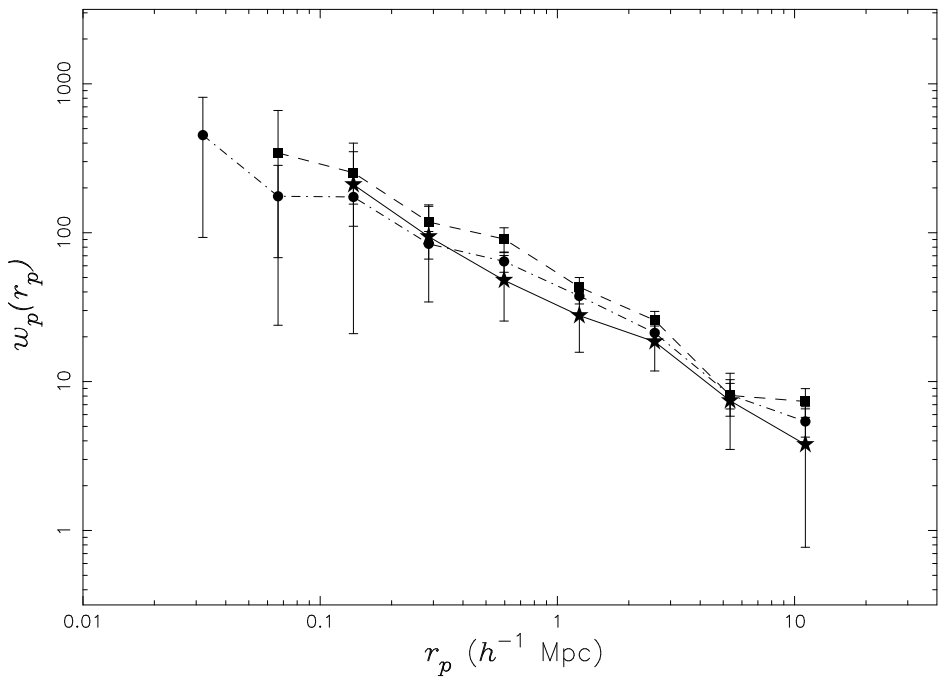}{3.00 in}{-90}{50}{50}{-198}{286}
\caption{
$w_p(r_p)$ divided by $W_0[$\ion{O}{2}] for $0.2 \le z \le 0.5$ for
the Abell 104 survey.  Galaxies with $W_0[$\ion{O}{2}] $> 10$\AA\ are
marked with filled stars, and galaxies with $W_0[$\ion{O}{2}] $< 10$\AA\
are marked with filled squares.  For comparison, $w_p(r_p)$ for the
whole sample is plotted with filled circles.}
\label{figures:A104_col_wp}
\end{figure}

\begin{figure}
\plotfiddle{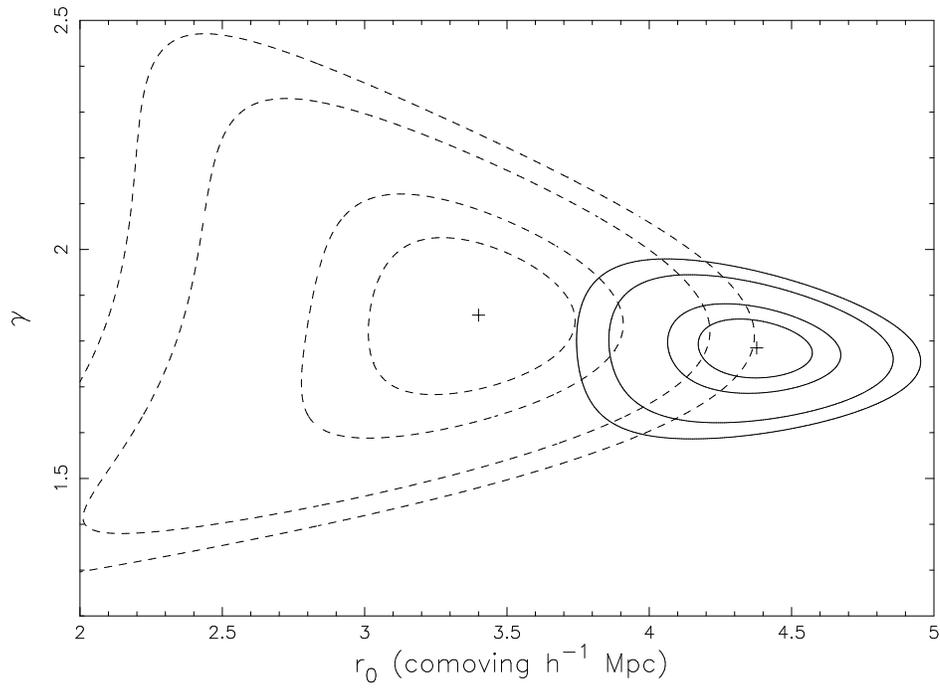}{3.00 in}{-90}{50}{50}{-198}{286}
\caption{
Error contours of $(r_0, \gamma)$ for power--law fits to the projected
two--point correlation functions of actively star--forming
($W_0[$\ion{O}{2}] $> 10$\AA\, dashed contours) and quiescent
($W_0[$\ion{O}{2}] $< 10$\AA\, solid contours) galaxies with $0.2 < z
< 0.5$ in the Abell 104 field.  The contour levels are the same as in
Figure \ref{figures:r0_gamma}.}
\label{figures:r0_gamma_col}
\end{figure}

In Figure \ref{figures:A104_col_wp}, we plot the projected correlation
functions of $0.2 \le z \le 0.5$ galaxies with $W_0[$\ion{O}{2}] $ >
10$\AA\ (filled stars) and with $W_0[$\ion{O}{2}] $ < 10$ \AA\ (filled
squares).  We also plot the projected correlation of the entire $0.2
\le z \le 0.5$ sample with filled circles.  As in local samples, the
galaxies with little or no ongoing star formation are more strongly
clustered than the blue, star--forming galaxies.  Error contours for
power--law fits to the projected correlation functions of the two
samples are shown in Figure \ref{figures:r0_gamma_col}.  While the
power--law indices ($\gamma$) of the two correlation functions are
indistinguishable within the errors, the correlation lengths differ at
the $2.4\sigma$ level, with the quiescent galaxies having a clustering
strength at $5 h^{-1}$ Mpc 1.6 times larger than that of the
star--forming galaxies.  The ratio of the correlation strengths is
comparable to that found in the Stromlo/APM redshift survey (1.7,
Loveday et al.\ 1995) but smaller than that found in the
Pisces--Perseus redshift survey (2.4, Guzzo et al.\ 1997), although
such comparisons are necessarily very rough because our division of
the $0.2 \le z \le 0.5$ sample by $W_0$[\ion{O}{2}] only approximates
division by morphological type.  Our data do not extend to high enough
redshift to test the claim of Le F\`evre et al.\ (1996) that there is
no difference in the clustering strengths of quiescent and
star--forming galaxies for $z \gtrsim 0.5$.  However, if future data
confirm Le F\`evre et al.'s claim, then the redshift interval over which the
clustering of quiescent galaxies has grown relative to star--forming
galaxies is $\Delta z \gtrsim 0.2$, or roughly one billion years.

\begin{figure}
\plotfiddle{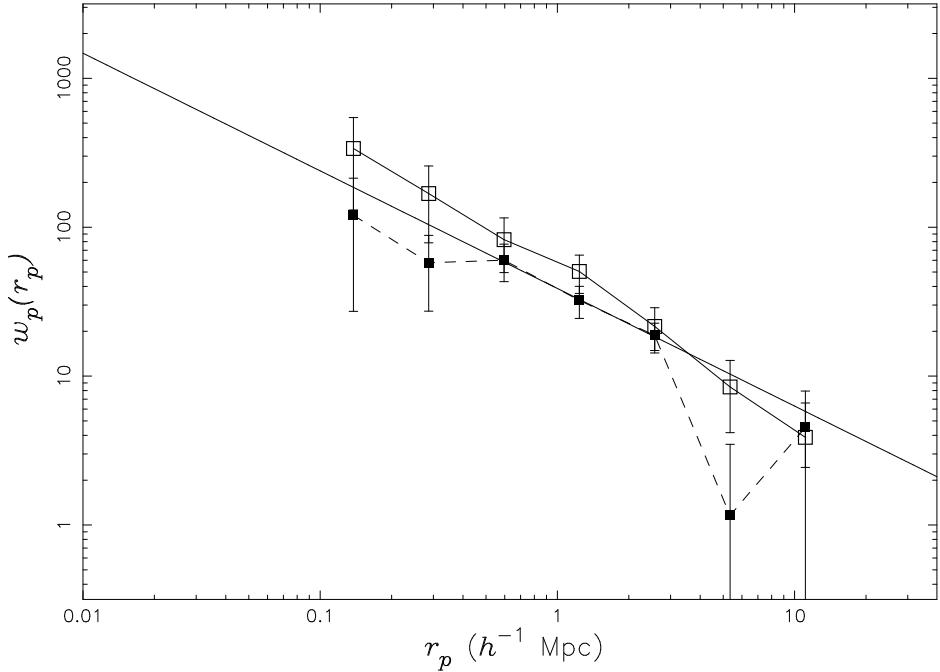}{3.00 in}{-90}{50}{50}{-198}{286}
\caption{
$w_p(r_p)$ divided by M($B_{AB}$) for $0.2 \le z \le 0.3$ for the
Abell 104 survey.  $w_p(r_p)$ for intrinsically luminous galaxies with
M($B_{AB}$) $\le -19.45 + 5\log h$ is plotted with filled squares, and
$w_p(r_p)$ for intrinsically faint galaxies with M($B_{AB}$) $\ge
-19.45 + 5\log h$ is plotted with filled stars.  The solid line shows
the best power--law fit to the projected correlation function of $0.2
\le z \le 0.3$ galaxies in the A104 field.  We have neglected the very
low point at $r_p = 5.3 h^{-1}$ Mpc when fitting the projected
correlation function of the intrinsically faint galaxies.}
\label{figures:A104_lum_wp}
\end{figure}

\begin{figure}
\plotfiddle{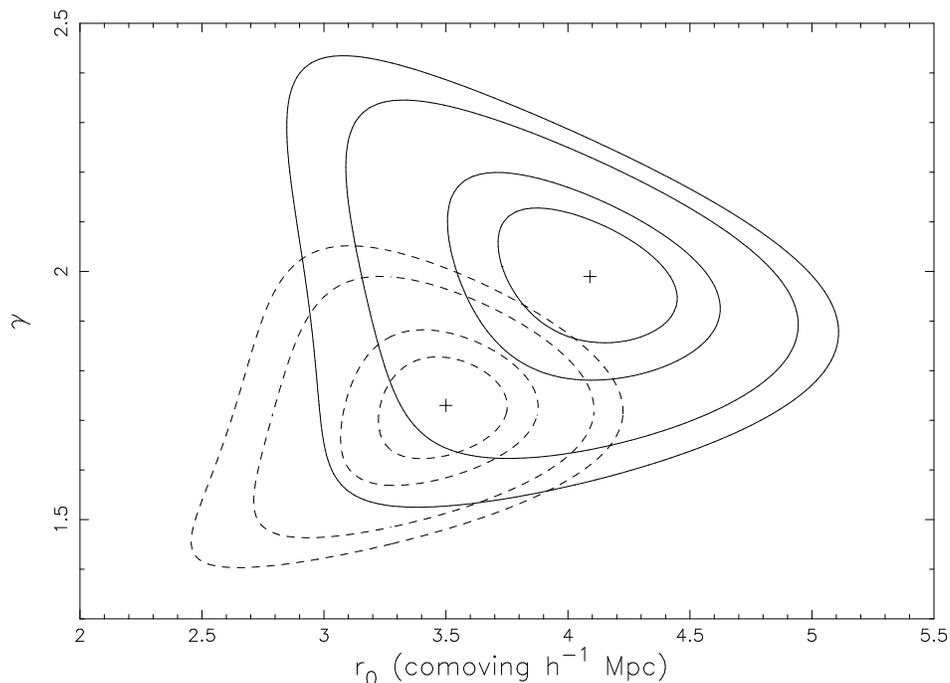}{3.00 in}{-90}{50}{50}{-198}{276}
\caption{
Error contours of $(r_0, \gamma)$ for power--law fits to the projected
two--point correlation functions of intrinsically faint (dashed contours) and
intrinsically bright (solid contours) galaxies with $0.2 < z < 0.3$ in the
Abell 104 field.  The contour levels are the same as in Figure
\ref{figures:r0_gamma}.}
\label{figures:r0_gamma_lum}
\end{figure}

As for division by star formation rate, the trends in clustering
observed for division by intrinsic luminosity at $0.2 \le z \le 0.3$
are similar to those observed at low redshift.  Since our galaxy
sample is selected by apparent magnitude, the range of absolute
magnitudes of galaxies in our sample varies with redshift.  We have
limited the redshift range to galaxies with $0.2 \le z \le 0.3$ to
avoid being biased towards super--$L^\ast$ galaxies at higher
redshift.  We divide the sample at $L^\ast$, M($B_{AB}$) $= -19.45 + 5
\log h$ for $q_0 = 0.5$ (Lin et al.\ 1998).  For $q_0 = 0.1$, we use
M($B_{AB}$) $= -19.58 + 5 \log h$ since a galaxy at $z = 0.30$ will
have an absolute magnitude 0.13 mag brighter in a $q_0 = 0.1$
cosmology than in a $q_0 = 0.5$ cosmology.  In Figure
\ref{figures:A104_lum_wp}, we plot the projected correlation functions
of galaxies brighter than and fainter than $L^\ast$ with filled
squares and filled stars, respectively.  The galaxies brighter than
$L^\ast$ appear to be more strongly clustered than the sub--$L^\ast$
galaxies and perhaps, as also seen in local studies, to have a steeper
correlation function slope.  These differences can be assessed
quantitatively in Figure \ref{figures:r0_gamma_lum}, where we plot
error contours of $(r_0, \gamma)$ for power--law fits to the projected
correlation functions.  The unusually low point at $r_p = 5.3 h^{-1}$
Mpc has been neglected in the fit for the intrinsically faint
galaxies.  The error contours reflect the visual impressions of the
differences between the two projected correlation functions plotted in
Figure \ref{figures:A104_lum_wp}.  However, neither the longer
correlation length nor the steeper slope of the correlation function
of the intrinsically luminous sample relative to the intrinsically
faint sample is statistically significant.

\section{The Pairwise Velocity Dispersion $\sigma_{12}$}

The redshift--space distortions of $\xi(r_p,r_\pi)$ contain
information on the velocity distribution function of galaxy pairs,
$P({\bf w} \vert {\bf r})$, where ${\bf w}$ is the velocity difference
of a pair with vector separation ${\bf r}$.  Peebles (1980) has
modeled $\xi(r_p,r_\pi)$ as a convolution of the real space
correlation function $\xi(r)$ with $P({\bf w} \vert {\bf r})$,
\begin{equation}
1 + \xi(r_p,r_\pi) = \int \lbrack 1+\xi(r)\rbrack P({\bf w} \vert {\bf r})
 d^3 {\bf w} \,.
\end{equation}
This expression can be simplified if we assume that the velocity
dispersion of pairs varies slowly with pair separation and that there
is no preferred direction in the velocity field.  With those
assumptions, $\xi(r_p,r_\pi)$ depends only on the distribution of
line--of--sight velocities, and we have
\begin{equation}
1 + \xi(r_p,r_\pi) = \int \lbrack 1+\xi(r)\rbrack P(v_{los} \vert r)
dv_{los}\,.
\end{equation}
If we separate ${\bf r}$ into real--space components $(r_p,y)$
perpendicular to and along the line--of--sight, then $r^2 = r^2_p +
y^2$, $v_{los} = H_0(r_\pi - y)$, and
\begin{equation}
1 + \xi(r_p,r_\pi) = \int \lbrack 1+\xi(\sqrt{r^2_p + y^2}) \rbrack
P(H_0(r_\pi - y) \vert r) dy \,.
\label{equations:model}
\end{equation}

It has been found in the analyses of previous surveys (Davis \&
Peebles 1983, Fisher et al.\ 1994, Marzke et al.\ 1995) that an
exponential distribution of pairwise line--of--sight velocities,
\begin{equation}
P(H_0(r_\pi - y) \vert r) = {1 \over {\sqrt{2}\sigma_{12}(r)}} 
\exp\left\{-\sqrt{2}H_0\left|{{r_\pi - y[1+v_{12}(r)/H_0r]} \over \sigma_{12}(r)}\right|\right\},
\end{equation}
where $v_{12}(r)$ is the mean relative velocity of galaxy pairs with
separation $r$ and $\sigma_{12}(r)$ is the pairwise velocity
dispersion along the line of sight, fits the data well.  The
exponential distribution also appears in $N$-body simulations (e.g.,
Zurek et al.\ 1994) and in theoretical analyses (Diaferio \& Geller
1996; Sheth 1996; Juszkiewicz, Fisher, \& Szapudi 1998).  We model
$v_{12}(r)$ using the streaming model of Davis \& Peebles (1983),
which is based on the similarity solution of the BBGKY equations,
\begin{equation}
v_{12}(r) = -H_0r {F \over {1 + (r/r_0)^2}}.
\end{equation}
Free expansion of pairs with Hubble flow corresponds to $F = 0$ (i.e.,
$v_{12}(r) = 0$), while stable clustering corresponds to $F = 1$.
This form matches results from $N$-body simulations modestly well
(Efstathiou et al.\ 1988, Zurek et al.\ 1994).  We are neglecting the
scale dependence of $\sigma_{12}$.  The Cosmic Virial Theorem (Peebles
1980) predicts that the dispersion of bound objects scales as
$\sigma_{12}(r) \propto r^{1-\gamma/2}$, which is only weakly
dependent on $r$ for $\gamma$ close to the observed value of $\approx
1.8$.

We estimate $\sigma_{12}$ by fitting Equation \ref{equations:model},
with $F$ held at 1, to the observed $\xi(r_p,r_\pi)$.  Since the
points of $\xi$ are correlated and the distribution of errors of $\xi$
is not Gaussian over portions of the $(r_p,r_\pi)$ plane (Fisher et
al. 1994), a traditional $\chi^2$--fitting procedure is not strictly
appropriate.  However, the off--diagonal elements of the covariance
matrix, computed using 50 bootstrap resamplings of the original
dataset, are typically 20 times smaller than the diagonal elements.
Since the error due to cosmic variance alone for our $2 \times 10^6
h^{-3}$ Mpc$^3$ ($q_0 = 0.5$) survey volume is expected to be
$\lesssim 20$\% (Marzke et al.\ 1995), we do not feel that an
elaborate analysis which accounts for the correlations in $\xi$ is
warranted.  We have thus used a straightforward $\chi^2$--fitting
procedure.

\begin{figure}
\plotfiddle{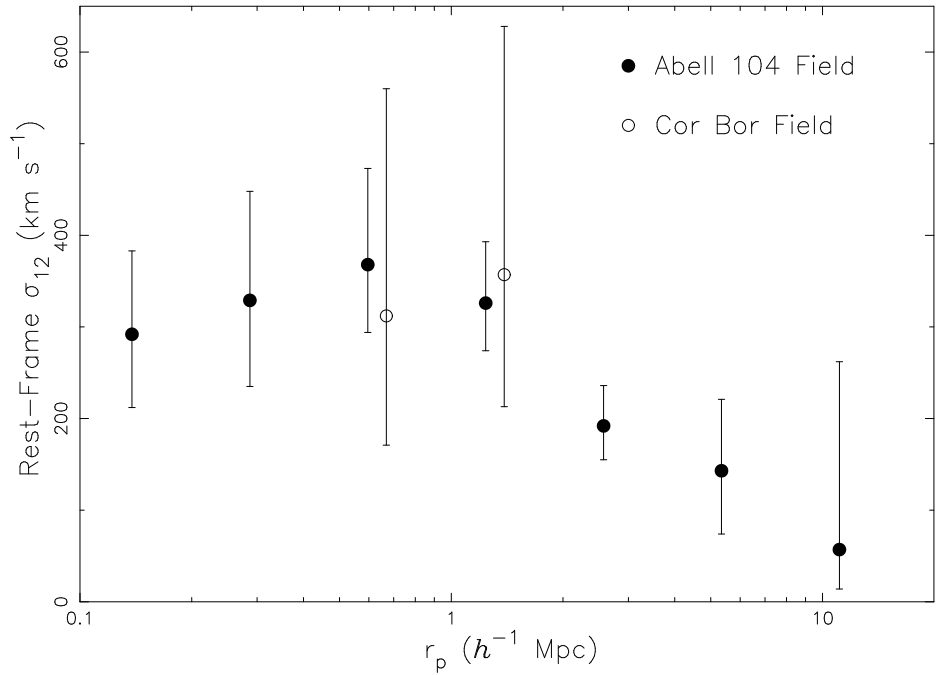}{3.00 in}{-90}{50}{50}{-198}{286}
\caption{
Rest--frame $\sigma_{12}$ for the A104 survey (filled circles) and the
Corona Borealis survey (unfilled circles, offset by 0.05 dex in $r_p$
for clarity) galaxies with $0.2 \le z \le 0.5$ ($z_{\rm med} = 0.30$)
as a function of projected separation $r_p$.  We have assumed that the
stable clustering hypothesis applies (i.e., $F \equiv 1$); our
estimates of of $\sigma_{12}$ are insensitive to $F$.}
\label{figures:sig12_r}
\end{figure}

The rest--frame values of $\sigma_{12}$ as a function of $r_p$ for
A104 and Corona Borealis galaxies with $0.2 \le z \le 0.5$ ($z_{\rm
med} = 0.30$) are summarized in Table 2 and plotted in Figure
\ref{figures:sig12_r}.  We obtain more precise estimates of
$\sigma_{12}$ from the A104 field since it contains significantly more
galaxies than the Corona Borealis field.  At $r_p = 1.24 h^{-1}$ Mpc,
the pairwise velocity dispersion in the A104 field is $\sigma_{12} =
326^{+67}_{-52}$ km s$^{-1}$.  Measurements of $\sigma_{12}$ at $r
\approx 1 h^{-1}$ Mpc for local samples range from $317^{+40}_{-49}$
km s$^{-1}$ for IRAS galaxies (Fisher et al.\ 1994) and
$345^{+95}_{-65}$ km s$^{-1}$ for late--type galaxies in the
Pisces--Perseus redshift survey (Guzzo et al.\ 1997) through $416 \pm
36$ km s$^{-1}$ for the Durham/UKST survey (Ratcliffe et al.\ 1998) to
$647 \pm 180$ km s$^{-1}$ for the combined CfA2+SSRS2 survey
(Marzke et al.\ 1995) and $570 \pm 80$ km s$^{-1}$ for the
Las Campanas survey (Jing, Mo, \& B\"{o}rner 1998).  Landy, Szalay, \&
Broadhurst (1998) have recently employed a novel Fourier decomposition
technique which naturally downweights the problematic contributions
from clusters of galaxies to estimate $\sigma_{12}$ for the Las
Campanas redshift survey.  They measure $\sigma_{12} = 363 \pm 44$ km
s$^{-1}$ for a survey with a total volume of $6 \times 10^6 h^{-3}$
Mpc$^3$; however, their neglect of coherent infall means that they
have probably underestimated $\sigma_{12}$ by $\sim 100$ km s$^{-1}$
(see Jing \& B\"{o}rner 1998).  Thus, our estimate of $\sigma_{12}$
for $0.2 \le z \le 0.5$ appears to be modestly lower than the values
measured locally.  Our estimate is also consistent with the
preliminary results from the CNOCII survey, $\sigma_{12} = 350 \pm 50$
km s$^{-1}$ for galaxies with $0.15 \le z \le 0.55$ (Carlberg et al.\
1999).

\section{Large--Scale Structure}

\begin{figure}
\plotfiddle{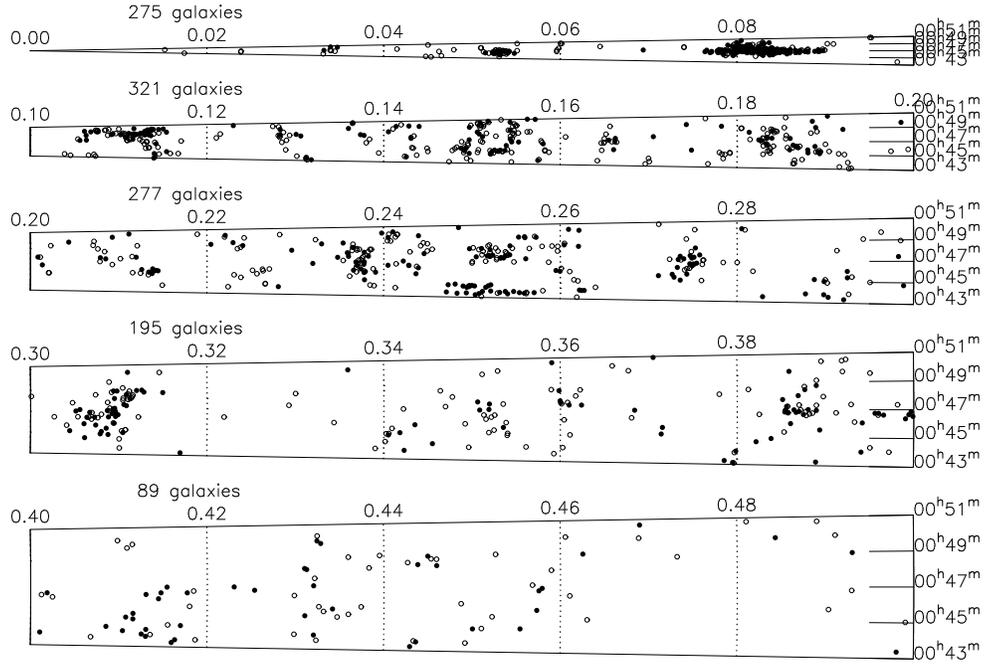}{3.00 in}{-90}{50}{50}{-198}{276}
\caption{
Redshift--Right--Ascension pie diagram for the A104 survey, split
into five panels of $\Delta z = 0.1$.  The filled and unfilled circles
represent galaxies dominated by an old stellar population and galaxies
dominated by star formation, respectively.  The prominent clump of
galaxies at $z \approx 0.08$ is Abell 104.  The comoving widths of the
panels at $z = 0.05, 0.15, 0.25, 0.35, {\rm and}\ 0.45$ are (for
$q_0 = 0.5$) 5.0, 14.1, 22.1, 29.2, and 35.5 $h^{-1}$ Mpc, respectively.}
\label{figures:A104_pie}
\end{figure}

The data in the A104 field provide a striking view of large--scale
structure out to $z = 0.5$.  The Corona Borealis data, because they
are much more sparsely sampled, are not as useful for exploring the
topology of the galaxy distribution.  The only comparable data in
terms of numbers of galaxies and depth in a similarly--sized
contiguous area are the data obtained by de Lapparent et al.\ (1997)
in the ESO--Sculptor Survey.  Those workers have obtained $\sim 700$
redshifts for $R \le 20.5$ mag galaxies in a $1.53^\circ \times
0.24^\circ$ region in the Sculptor constellation and describe patterns
in the large--scale galaxy distribution similar to those we report
here.  Our $r \le 21$ mag data are plotted in a
redshift--right--ascension ``pie'' diagram in Figure
\ref{figures:A104_pie}.  In order to fit on one page, we have split
the diagram into five segments, each of which has a length of 0.1 in
redshift space.  The number of galaxies with measured redshifts in
each segment is listed above the segment, and the galaxies are plotted
with different symbols according to their gross spectral properties.
Galaxies represented by solid circles have spectra dominated by an old
stellar population and have weak or no visible emission lines.
Galaxies represented by unfilled circles are actively forming stars
and have easily visible emission lines.  The structure delineated by
the galaxies is strongly reminiscent of the structure seen in local
redshift surveys (e.g., Geller \& Huchra 1989), namely, nearly all the
galaxies lie in larger structures (typically $\sim50 - 60 h^{-1}$
Mpc), walls and bubbles, which bound large empty regions.  This
diagram reveals that the structure seen in the local universe
continues out to at least $z = 0.5$.  The visual impression is that
only a small fraction of galaxies ($\lesssim 10$\%) are isolated field
galaxies.

As an attempt to quantify the structure, we have performed a
friends--of--friends analysis of the A104 field.  A
friends--of--friends analysis is a standard means to isolate clumps of
galaxies at a given overdensity (Huchra \& Geller 1982, Nolthenius \&
White 1987).  The overdensity level is specified by the linking
parameter, $l$, which is made dimensionless by scaling it by the
average separation of galaxies (as a function of redshift).  The
analysis of an observed galaxy catalog is complicated by the fact that
peculiar velocities distort large scale structure along the line of
sight.  In particular, dense clumps of galaxies, such as galaxy
clusters, have large velocity dispersions and appear elongated and
less dense in redshift space.  Since the redshift--space map is
undistorted perpendicular to the line of sight, we use separate
linking parameters along the line of sight and perpendicular to the
line of sight.  For simplicity, we use redshift--independent linking
velocities, $l_{\rm z}$, of either 350 or 500 km s$^{-1}$ along the
line of sight, the first value being approximately equal to the
pairwise velocity dispersion and the second value being a good
compromise to prevent the ``fingers--of--God'' from being split off
from clusters while still preventing an obvious overmerging of
structures.  We have run the friends--of--friends analysis with three
different transverse linking parameters, $l_{\rm trans} = $ 0.2, 0.5,
and 0.7.  Since the overdensity of identified clusters scales roughly
as $\delta \rho / \rho \sim 2/l^3$, these linking parameters select
structures with overdensities of roughly 250, 16, and 5, corresponding
respectively to virialized structures, collapsing but not yet
virialized structures, and structures just reaching turnaround and
starting to recollapse.  An additional complication is that the
observed galaxy density in our sample declines with redshift and
varies with spatial position.  We therefore scale the transverse
linking length as $[s_{ij}\bar n_{gal}(z)]^{-1/3}$, where $n_{gal}(z)$
is the mean observed galaxy density as a function of redshift assuming
a strict $r \le 21$ apparent magnitude limit and $s_{ij}$ is the
average of the values of the total selection function $S(m, \alpha,
\delta)$ at the positions of galaxies $i$ and $j$.

\begin{figure}
\plotfiddle{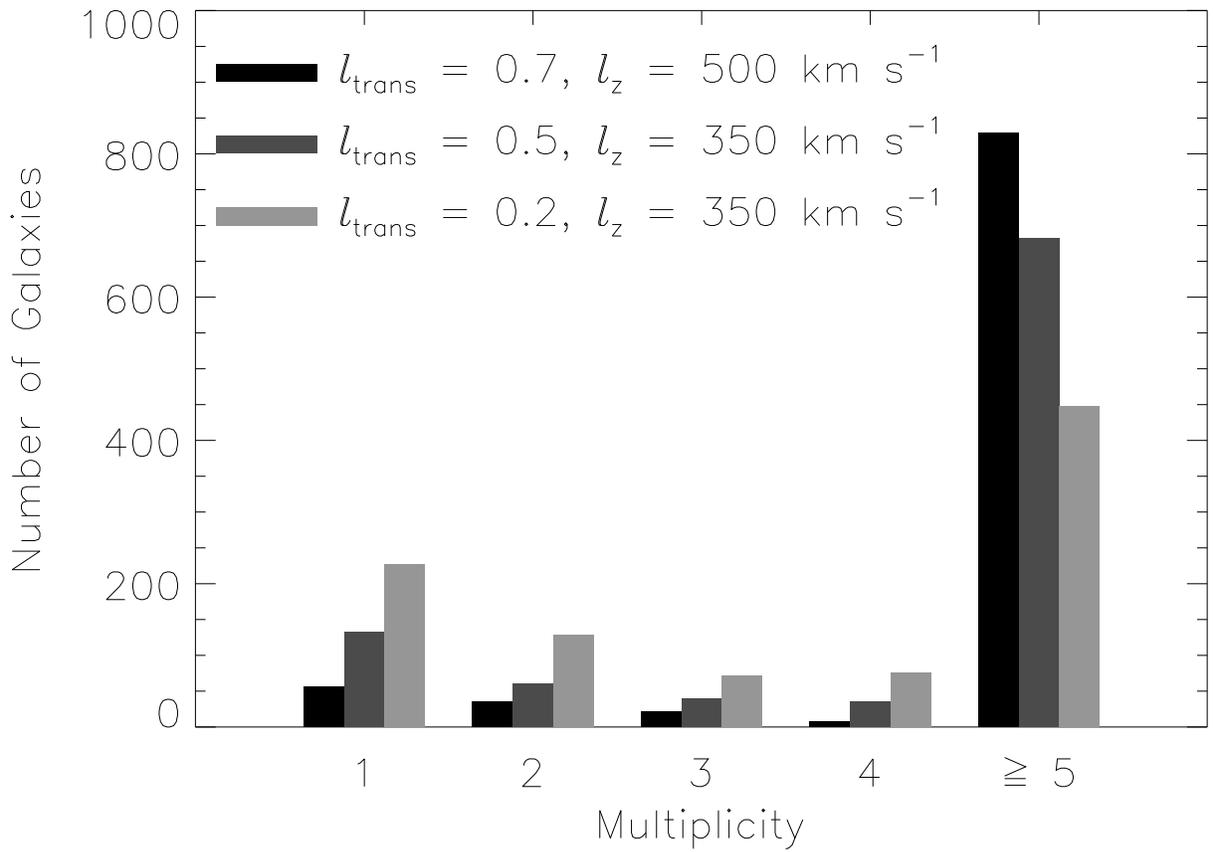}{5.00 in}{0}{100}{100}{-305}{-376}
\caption{
Number of A104 survey field galaxies ({\it excluding} galaxies in the
A104 cluster itself) as a function of the number of group members for
friends--of--friends analyses with three pairs of parameters:  dark
gray, $l_{\rm trans} = 0.7$, $l_{\rm z} = 500$ km s$^{-1}$; medium
gray, $l_{\rm trans} = 0.5$, $l_{\rm z} = 350$ km s$^{-1}$; and light
gray, $l_{\rm trans} = 0.2$, $l_{\rm z} = 350$ km s$^{-1}$).}
\label{figures:A104_fof}
\end{figure}

The results of our friends--of--friends analysis for the A104 survey
field, with A104 itself {\it excluded}, are plotted in Figure
\ref{figures:A104_fof}.  For three different combinations of
line--of--sight and transverse linking parameters, we show the number
of galaxies in groups of multiplicity 1 (i.e., isolated galaxies), 2,
3, 4, and $\ge 5$.  As expected, the general trend in the plot is
that, as the density threshold is raised, fewer galaxies reside in
large clumps and more galaxies are found in small groups or are
entirely isolated.  It is also striking, however, how large a fraction
of galaxies are part of groups with five or more members.  At our
lowest overdensity threshold, $\sim 5$, 87\% of galaxies have five or
more linked companions, and this fraction only declines to
approximately 50\% at an overdensity threshold of $\sim 250$.  It is
important to interpret these numbers cautiously due to the sensitivity
of the results to the precise value of the linking parameter, the
uncertainties involved with the redshift--space distortions, and the
only approximate relationship between the linking parameter and
overdensity threshold.  Nevertheless, these results do support our
visual impressions of the galaxy distribution displayed in Figure
\ref{figures:A104_pie}.  As a further illustration that the
friends--of--friends of analysis is identifying credible galaxy
structures (at an overdensity of $\sim 5$), we replot in Figures
\ref{figures:A104_fof_ra} and \ref{figures:A104_fof_dec}, which are
projections over declination and right ascension, respectively, the
galaxy distribution shown in Figure \ref{figures:A104_pie} but with
all the galaxies in a given group marked in the same color.  Aside
from occasional illusions caused by the projections over one
dimension, there is no doubt that the friends--of--friends analysis is
finding plausible galaxy structures.

We have also applied a friends--of--friends analysis to dark matter
halos identified in $N$--body simulations and will discuss the results
in \S 7.2.

\section{Discussion and Conclusions}

We have presented an analysis of the clustering, pairwise velocity
dispersion, and large--scale structure in two independent redshift
surveys of field galaxies at intermediate redshifts ($0.2 < z < 0.5$).
Our combined survey sparsely covers a very large region ($\sim 20$
sq.\ deg.) and includes redshifts for 835 galaxies with $r \le 21$ mag
and $0.2 < z < 0.5$.  The large area reduces the errors due to cosmic
variance down to levels comparable to the statistical errors (i.e., at
$1 h^{-1}$ Mpc, $\lesssim 10$\% for the correlation strength and
$\lesssim 20$\% for the pairwise velocity dispersion), providing a
firm foundation for analyzing the evolution of clustering and the
pairwise velocity dispersion.

\subsection{Redshift Evolution of $\xi$ and $\sigma_{12}$}

Assuming that $\xi(r,z)$ is well fit by a power-law
$(r/r_0)^{-\gamma}$, it has been customary to parameterize the evolution
of $\xi(r,z)$ with a power law (Groth \& Peebles 1977):
\begin{equation}
\xi(r,z) = \xi(r,0)(1+z)^{-(3+\epsilon-\gamma)}\,,
\label{equations:epsilon}
\end{equation}
where $r$ is the comoving separation.
For clustering which is fixed in comoving coordinates, the
evolutionary parameter $\epsilon = \gamma-3$.  For clustering which is
fixed in physical coordinates, $\epsilon = 0$.  Linear theory applied
to the matter distribution predicts $\epsilon = \gamma-1$ (Peebles
1980).  The comoving correlation length can be written as 
\begin{equation}
r_0(z) = r_0(z = 0) (1+z)^{-(3+\epsilon-\gamma)/\gamma}.
\label{equations:r_0}
\end{equation}

It is likely that this prescription is too simplistic.  First, the
linear theory prescription only applies to the mass distribution.  In
cold dark matter scenarios, galaxies form in dark matter halos, and so
one should compare the measured clustering of galaxies with the
clustering of halos.  Second, galaxies may form in halos only in a
biased fashion, and this bias may change with redshift.  $N$-body
simulations demonstrate that the Groth \& Peebles model is not
accurate (Colin et al.\ 1997; Ma 1999).  The evolution of $\xi$ for
halos is substantially slower that the evolution of $\xi$ for matter,
and its redshift dependence is {\it not} well--described by the
power-law model given in Equation \ref{equations:epsilon}.

In Figure \ref{figures:r0_evol}, we plot an assortment of measurements
of the comoving correlation length as a function of redshift.  Low
redshift data come from the combined CfA2/SSRS2 survey (Marzke et al.\
1995), the Stromlo/APM survey (Loveday et al.\ 1995), the Las Campanas
Redshift Survey (Jing et al. 1998), and the infrared--selected 1.2 Jy
IRAS survey (Fisher et al.\ 1994).  At high redshifts, we plot the
measurements presented here, the $z = 0.37$ point from the CNOCI
survey (Shepherd et al.\ 1997); the $z = 0.08$, $z = 0.14$, $z =
0.28$, and $z = 0.43$ points from the CNOCII survey (Carlberg et al.\
1999); the $z = 0.34$, $z = 0.62$, and $z = 0.86$ points from the CFRS
survey (Le F\`evre et al.\ 1996); and the $z = 0.5$ estimate obtained
by Postman et al.\ (1998) by deprojecting their wide--field $I$-band
imaging data.  For ease of comparison with previous work, the points
plotted here are for $q_0 = 0.1$ (or $q_0 = 0$).  Note that the three
solid squares marking our measurements for the Abell 104 field are not
independent.  The low redshift point is for galaxies in the redshift
interval $0.2 < z <0.3$, the high redshift point is for galaxies in
the redshift interval $0.3 < z < 0.5$, and the point in between is for
galaxies in the combined redshift interval $0.2 < z < 0.5$.  Although
there is a wide dispersion in the measurement of $r_0$ at all
redshifts, there is a strong indication that the comoving correlation
length declines slowly beyond $z\approx 0$.  The pioneering CFRS
survey suggests the most dramatic decline in the correlation length.
The CFRS results, however, should be interpreted with caution since
the survey covered such a small volume and includes mainly
sub--$L^\ast$ galaxies at $z \lesssim 0.5$.  The CNOCI measurement is
also low, but it too may suffer from cosmic variance caused by
analyzing only a small volume.  The results from the larger volumes
surveyed by CNOCII, Postman et al.\ (1998), and ourselves are all
consistently above the CFRS and CNOCI points and suggest a more modest
decline in the comoving correlation length with redshift.

As an illustration, we also plot in Figure \ref{figures:r0_evol}
estimates of the comoving correlation length for dark matter halos
with masses greater than $2.5 \times 10^{12}$ M$_\odot$ from the
$N$--body simulations presented in Ma (1999).  Results for three
cosmological models are shown: the standard $\Omega_m=1$ cold dark
matter model with $\sigma_8 = 0.7$ and two low-density models with a
cosmological constant, $(\Omega_m,\Omega_\Lambda) = (0.5,0.5)$ and
(0.3, 0.7), normalized to {\it COBE}.  For each model, we have only
plotted the correlation length at one redshift because the variation
of the comoving correlation length is less than 10\% over the entire
redshift range from $z = 0$ to $z = 1.5$.  The correlation length at
$z \sim 0.4$ has a clear dependence on the matter density parameter
$\Omega_m$.  The models with lower $\Omega_m$ have larger $r_0$
because the universe is dominated by the vacuum energy at this low
redshift and gravitational clustering has effectively stopped.
However, it should be remembered that the correlation length depends
on the halo population, and there may exist a nontrivial relationship
between the dark matter halo distribution in simulations and the
observed galaxy distribution.  Recent work has proposed ways to make
connections between halo and galaxy distributions by combining
semi--analytic models of galaxy evolution with traditional $N$-body
simulations (Kauffmann et al.\ 1999, 1998; Baugh et al.\ 1998).  While the
uncertainties associated with these semi--analytic models are still
large, the models can produce plausible galaxy evolution histories and
can match many observed properties of the evolving galaxy population
from $z \sim 3$ to the present.  The details of the predicted
clustering evolution depend on a large number of factors, including
the cosmological model, the nature of the dark matter, the type of
galaxies observed, the waveband of the observations, and so on.  The
sensitive dependence of correlation functions on sample selection
obviates---for the time being---the use of measurements of the
evolution of clustering to constrain cosmological parameters.
However, as emphasized by Kauffmann et al.\ (1998), this same
sensitivity makes studies of clustering evolution a powerful tool for
constraining models of galaxy formation and evolution.

Given the uncertainties in the observations and model predictions, it
appears premature to attempt to draw any firm conclusions about
cosmological parameters or galaxy evolution from the data presented in
Figure \ref{figures:r0_evol}.  It is nonetheless worthwhile to
highlight the general agreement between the data and the predictions
of hierarchical structure formation models.  While the clustering of
the underlying matter distribution declines monotonically with
increasing redshift, the comoving correlation length of galaxies is
predicted to decline modestly or, in fact, remain flat until $z \sim
1.5$ and then rise at very high redshifts as the only galaxies bright
enough to be observable with current instrumentation become very
highly biased (Kauffmann et al.\ 1998, Baugh et al.\ 1998).  The data
collected in Figure \ref{figures:r0_evol}, excluding the unusually low
results from the CFRS and CNOCI surveys, show only a $1 - 2 h^{-1}$
Mpc decline in comoving correlation length to $z \sim 0.5$.  At high
redshifts beyond the redshift range plotted in Figure
\ref{figures:r0_evol}, the correlation length of $z \sim 3$
Lyman--break galaxies is roughly comparable to that of local $L^\ast$
galaxies (Steidel et al.\ 1998, Giavalisco et al.\ 1998, Adelberger et
al.\ 1998).  Thus, the observed evolution of the galaxy correlation
function broadly matches that expected in hierarchical structure
formation models.  Furthermore, the observed variation of the
correlation length with galaxy population at a given redshift is in
accord with expectations from semi--analytic models.  For example,
Kauffmann et al.'s models naturally explain the weaker clustering of
blue and less luminous galaxies relative to red and more luminous
galaxies as reflections of the typical masses of the dark halos in
which different galaxies reside (i.e., blue and less luminous galaxies
reside in less massive, and therefore less biased, halos than red and
more luminous galaxies).  As both the observations and the theoretical
models are refined, it is clear that studies of the evolution of
galaxy clustering will provide valuable insights into galaxy formation
and evolution.

As discussed in \S 5 above, the pairwise velocity dispersion
$\sigma_{12}$ is a more difficult quantity to measure reliably than
the spatial two--point correlation function $\xi$.  Our measurement of
$\sigma_{12}(r_p = 1.24 h^{-1} \rm{Mpc}) = 326^{+67}_{-52}$ km
s$^{-1}$ at $z_{med} = 0.3$ indicates a $\sim 25$\% decrease from the
typical values measured locally, although our value is comparable to
the pairwise velocity dispersion of local samples selected in the
infrared or with early--type galaxies or clusters excluded.
Preliminary results from CNOCII are consistent with our finding.

\subsection{Very Large Scale Structure}

\begin{figure}
\plotfiddle{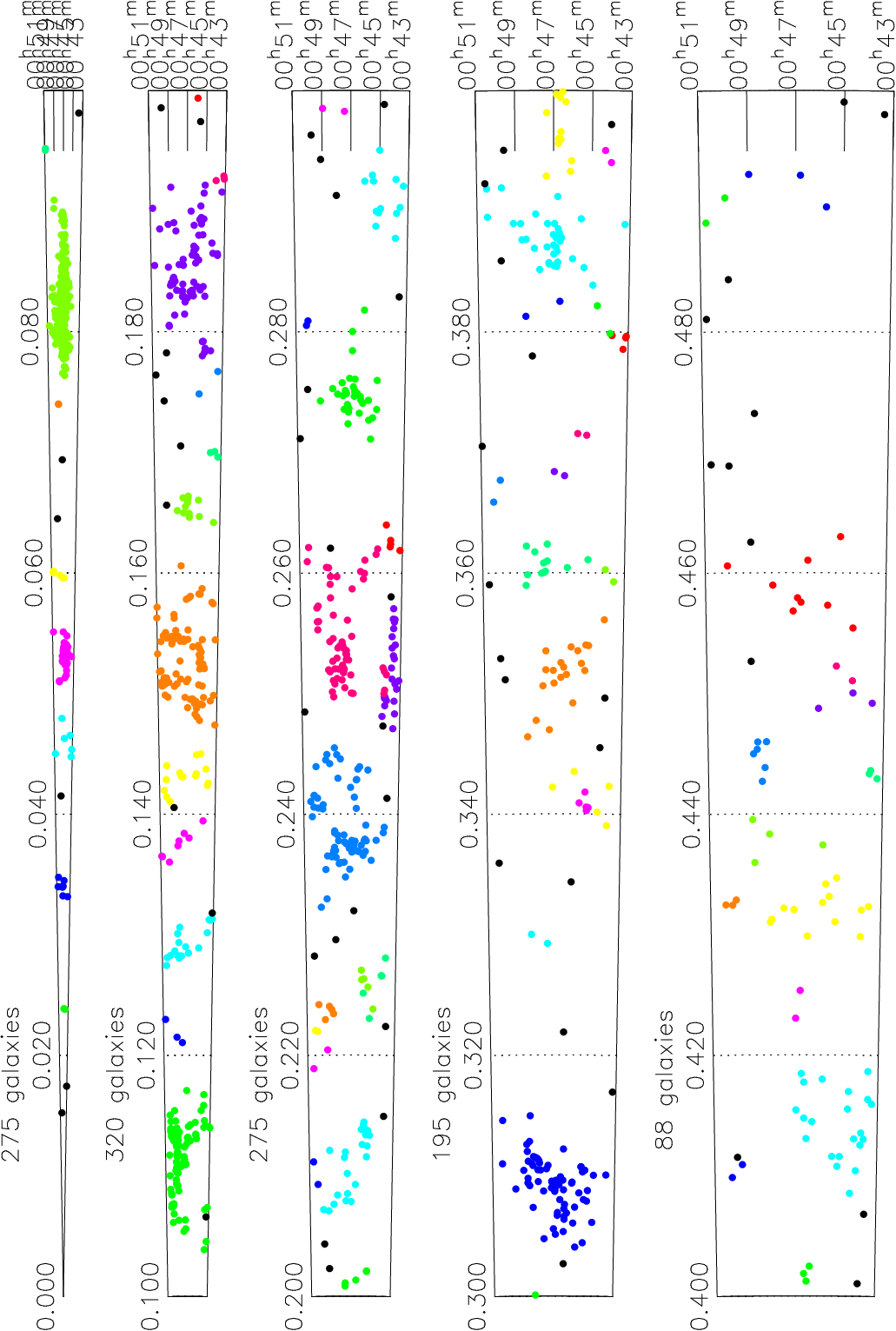}{3.00 in}{-90}{50}{50}{-198}{276}
\caption{
Redshift--Right--Ascension pie diagram for the A104 survey, split into
five panels of $\Delta z = 0.1$, with friends--of--friends selected
structures with $\delta \rho / \rho \sim 5$ ($l_{\rm trans} = 0.7$,
$l_{\rm z} = 500$ km s$^{-1}$) marked by the sequence of colors.  The
comoving widths of the panels at $z = 0.05, 0.15, 0.25, 0.35, {\rm
and}\ 0.45$ are (for $q_0 = 0.5$) 5.0, 14.1, 22.1, 29.2, and 35.5
$h^{-1}$ Mpc, respectively.}
\label{figures:A104_fof_ra}
\end{figure}

\begin{figure}
\plotfiddle{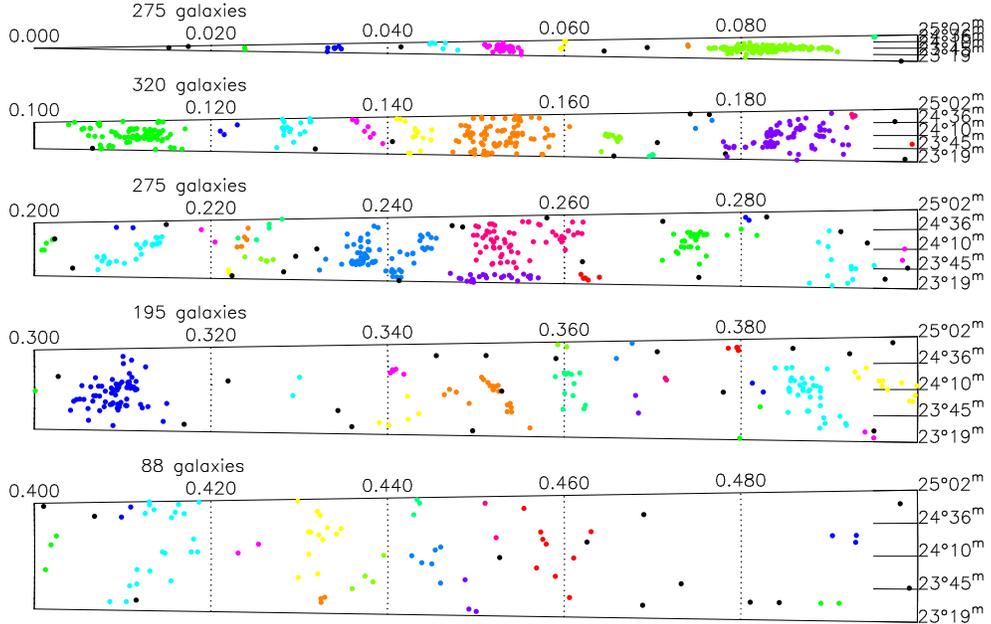}{3.00 in}{-90}{50}{50}{-198}{260}
\caption{
Redshift--Declination pie diagram for the A104 survey, split into five
panels of $\Delta z = 0.1$, with friends--of--friends selected
structures with $\delta \rho / \rho \sim 5$ ($l_{\rm trans} = 0.7$,
$l_{\rm z} = 500$ km s$^{-1}$) marked by the sequence of colors.  The
comoving widths of the panels at $z = 0.05, 0.15, 0.25, 0.35, {\rm
and}\ 0.45$ are (for $q_0 = 0.5$) 5.0, 14.1, 22.1, 29.2, and 35.5
$h^{-1}$ Mpc, respectively.}
\label{figures:A104_fof_dec}
\end{figure}

\begin{figure}
\plotfiddle{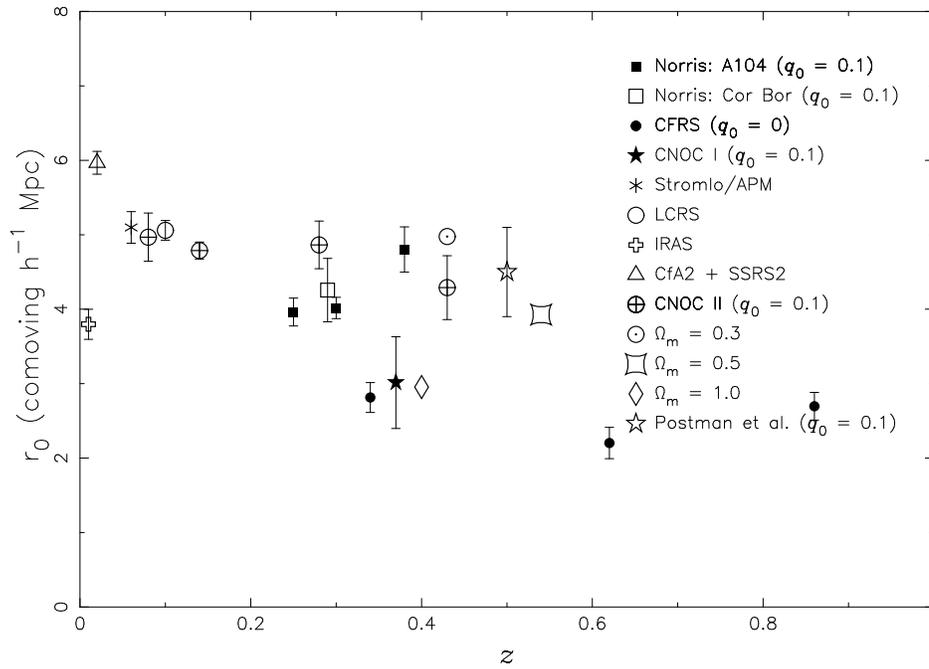}{3.00 in}{-90}{50}{50}{-198}{276}
\caption{ 
The comoving correlation length $r_{0,\rm comoving}$ as a
function of redshift for a heterogeneous assortment of redshift and
imaging surveys.  The higher redshift points are all plotted for $q_0
= 0.1$ (or $q_0 = 0$).  Also plotted are comoving correlation lengths for
halos with masses larger than $2.5 \times 10^{12}$ M$_\odot$
identified in the $N$--body simulations of Ma (1999).  See text
and plot legend for details.}
\label{figures:r0_evol}
\end{figure}

The map of the large--scale galaxy distribution out to $z = 0.5$ in
the Abell 104 field presented in Figures \ref{figures:A104_fof_ra} and
\ref{figures:A104_fof_dec} exhibits a striking structure in which the
galaxies lie mainly in thin sheets surrounding large ($\sim 50 - 60
h^{-1}$ Mpc), nearly empty voids, similar to the pattern of structure
seen in local surveys.  Using a friends--of--friends analysis, we
found that only $\sim 10$\% of galaxies in the survey did not lie in a
structure with an overdensity of at least $\sim 5$.  We have also
performed a {\it real}--space friends--of--friends analysis of the
halos formed in the $N$--body simulations described in the previous
section.  Although the precise numbers depend very sensitively on the
linking parameter chosen and the minimum mass of the selected halos,
the results from the $N$--body simulations are compatible with the
observations.  For example, the percentage of halos with masses
greater than $10^{12}$ M$_\odot$ which are part of larger structures
with overdensities $\gtrsim 5$ ($l \approx 0.7$) is approximately 80\%
for the standard cold dark matter model and 70\% for the $(\Omega_m,
\Omega_\Lambda) = (0.3, 0.7)$ model.  As an illustration of the
sensitivity of these numbers to the chosen parameters, we note that
100\% of the $10^{12}$ M$_\odot$ and greater halos are linked together
in the low--density model when the linking parameter is raised to $l
\approx 0.9$.

The pattern of sheets and shells visible in the galaxy distribution,
if fortuitously aligned along the line of sight, could be responsible
for the apparent $128 h^{-1}$ Mpc periodicity observed by Broadhurst
et al.\ (1990) in their pencil--beam survey of the North and South
Galactic poles.  This pattern is probably also reflected in the
prominent redshift--space spikes observed by Cohen et al.\ (1996) in
their K--band--selected redshift survey.  However, as emphasized by
Kaiser \& Peacock (1991), the appearance of periodic structures in
pencil--beam redshifts surveys is exaggerated by projection of
small--scale power in the three--dimensional power spectrum to large
scales in the one--dimensional power spectrum.  Nevertheless, there
remain credible detections of excess power on $\sim 100h^{-1}$ Mpc
scales, particularly in the two--dimensional power spectrum of
galaxies in the Las Campanas Redshift Survey (Landy et al.\ 1996) and
in the distibution of rich clusters of galaxies (Einasto 1998).  We
are, therefore, continuing our survey in the Corona Borealis region in
order to be able to compute the three--dimensional power spectrum at
$z_{med} \approx 0.3$ and to delineate structure on {\it transverse}
scales of $\sim 100 h^{-1}$ Mpc at $z \sim 0.5$.  Additional data
will, of course, also allow improved estimates of the
intermediate--redshift correlation function and pairwise velocity
dispersion.

\acknowledgements

We are grateful to the Kenneth T. and Eileen L. Norris Foundation for
their generous grant for construction of the Norris Spectrograph.  We
thank the staff of the Palomar Observatory for the expert assistance
we have received during the course of our Norris surveys, Ray Carlberg
for helpful comments and encouragement, and the referee, Ron Marzke,
for suggestions to improve and clarify the presentation.  CPM
acknowledges a Penn Research Foundation Award, and WLWS acknowledges
NSF Grant AST-9529093.

\begin{deluxetable}{ccccccc}
\tablecolumns{7}
\tablewidth{0pt}
\tablenum{1}
\tablecaption{Best--Fit Power--Law Correlation Function Parameters}
\tablehead{
\colhead{Sample} &
\colhead{Redshift Range} &
\colhead{$N_{\rm gal}$} &
\colhead{$z_{\rm med}$} &
\multicolumn{2}{c}{$r_0$ ($h^{-1}$ Mpc)\tablenotemark{a}} &
\colhead{$\gamma$} \\[0.2ex]
\colhead{} &
\colhead{} &
\colhead{} &
\colhead{} &
\colhead{$q_0 = 0.1$} &
\colhead{$q_0 = 0.5$} &
\colhead{}}
\startdata
\sidehead{\hfil Abell 104 Survey Field \hfil}
all galaxies 			& $0.2 < z \le 0.5$ & 558 & 0.30 
	& $4.02 \pm 0.14$ & $3.70 \pm 0.13$ & $1.77 \pm 0.05$ \nl
$W_0[$\ion{O}{2}] $<$ 10\AA 	& $0.2 < z \le 0.5$ & 349 & 0.31 
	& $4.46 \pm 0.19$ & $4.16 \pm 0.18$ & $1.84 \pm 0.07$ \nl
$W_0[$\ion{O}{2}] $>$ 10\AA 	& $0.2 < z \le 0.5$ & 201 & 0.29 
	& $3.80 \pm 0.40$ & $3.38 \pm 0.37$ & $1.85 \pm 0.18$ \nl
all galaxies			& $0.2 < z \le 0.3$ & 275 & 0.25
	& $3.96 \pm 0.19$ & $3.70 \pm 0.19$ & $1.75 \pm 0.06$ \nl
$M(B_{AB}) < M^\ast$\tablenotemark{b} 	& $0.2 < z \le 0.3$ & 99  & 0.25 
	& $4.34 \pm 0.38$ & $4.09 \pm 0.37$ & $1.99 \pm 0.14$ \nl
$M(B_{AB}) > M^\ast$\tablenotemark{b} 	& $0.2 < z \le 0.3$ & 175 & 0.25 
	& $3.84 \pm 0.27$ & $3.50 \pm 0.27$ & $1.73 \pm 0.10$ \nl
all galaxies			& $0.3 < z \le 0.5$ & 283 & 0.38
	& $4.80 \pm 0.30$ & $4.26 \pm 0.25$ & $1.68 \pm 0.08$ \nl
all galaxies			& $0.32 < z \le 0.5$ & 212 & 0.39
	& $4.16 \pm 0.35$ & $3.78 \pm 0.32$ & $1.72 \pm 0.12$ \nl
\sidehead{\hfil Corona Borealis Survey Field \hfil}
all galaxies			& $0.2 < z \le 0.5$ & 277 &
0.29
	& $4.26 \pm 0.42$ & $3.92 \pm 0.35$ & $1.63 \pm 0.10$ \nl
\enddata
\tablenotetext{a}{comoving}
\tablenotetext{b}{using ($h = 1$) $M^\ast = -19.58,\ -19.45$ for $q_0 = 0.1,\ 0.5$} 
\end{deluxetable}

\begin{deluxetable}{ccc}
\tablecolumns{3}
\tablewidth{0pt}
\tablenum{2}
\tablecaption{Rest--Frame Pairwise Velocity Dispersion at $z_{\rm med}
= 0.30$}
\tablehead{
\colhead{} &
\colhead{Abell 104 Field} &
\colhead{Corona Borealis Field} \\[0.2ex]
\colhead{$r_p$} &
\colhead{$\sigma_{12}$} &
\colhead{$\sigma_{12}$} \\[0.2ex]
\colhead{($h^{-1}$ Mpc)} &
\colhead{(km s$^{-1}$)} &
\colhead{(km s$^{-1}$)}}
\startdata
 0.14 & $292^{+91}_{-80}$  & \nodata  \nl
 0.29 & $329^{+119}_{-94}$ & \nodata \nl
 0.60 & $368^{+105}_{-74}$ & $312^{+248}_{-141}$ \nl
 1.24 & $326^{+67}_{-52}$  & $357^{+271}_{-144}$ \nl
 2.57 & $192^{+44}_{-37}$  & \nodata \nl
 5.34 & $143^{+78}_{-69}$  & \nodata \nl
11.11 &  $57^{+205}_{-43}$ & \nodata \nl
\enddata
\tablecomments{All fits hold $F \equiv 1$, which corresponds to
assuming that the stable clustering hypothesis is correct and applies
on the scales probed here.  Our estimates of $\sigma_{12}$ only weakly
depend on this assumption.}
\end{deluxetable}

\newpage

\clearpage

\end{document}